\documentclass[12pt]{article}
\usepackage{amssymb,amsmath,epsfig}
\usepackage{graphicx}
\allowdisplaybreaks

\begin{document}

\title{\bf Extended Gravitational Decoupled Solutions in Self-interacting Brans-Dicke Theory}

\author{M. Sharif \thanks{msharif.math@pu.edu.pk} and Amal Majid
\thanks{amalmajid89@gmail.com}\\
Department of Mathematics, University of the Punjab,\\
Quaid-e-Azam Campus, Lahore-54590, Pakistan.}

\date{}

\maketitle
\begin{abstract}
In this paper, we construct anisotropic spherical solutions from
known isotropic solutions through extended gravitational decoupling
method in the background of self-interacting Brans-Dicke theory. The
field equations are decoupled into two sets by applying geometric
deformations on radial as well as temporal metric components. The
first array corresponds to isotropic fluid while influence of the
anisotropic source is confined to the second set. The isotropic
sector is determined through metric functions of isotropic solutions
(Tolman IV/Krori-Barua) whereas two constraints on the anisotropic
source are required to close the second system. The impact of the
massive scalar field as well as the decoupling parameter on the
physical characteristics of the anisotropic solutions is analyzed
graphically. We also check the viability, compactness, surface
redshift and stability of the obtained solutions. It is found that
the resulting solutions follow accepted physical trend for some
values of the decoupling parameter.
\end{abstract}
{\bf Keywords:} Brans-Dicke theory; Gravitational decoupling;
Self-gravitating systems.\\
{\bf PACS:} 04.50.Kd; 04.40.-b; 04.40.Dg

\section{Introduction}

The universe is a well-structured yet incomprehensible system
composed of heavenly bodies and other mysterious components. The key
to understand the evolution of the vast cosmos lies in the study of
the arrangement as well as the physical behavior of celestial
objects. In this regard, general relativity (GR) played a remarkable
role in providing elementary insights to the mechanism governing the
interior of astronomical bodies. The exact solutions of the
non-linear field equations describe the intricate nature of
relativistic objects. Schwarzschild \cite{1}, the pioneer, obtained
a solution describing a spherical object with an incompressible
perfect fluid in its interior. Many researchers followed suit and
constructed more interior solutions. However, the non-linearity of
the equations poses a hindrance to extracting physically realistic
solutions.

Lemaitre \cite{1'} observed that anisotropy occurs in low as well as
high-density profiles due to rotational motion, phase transition, or
presence of magnetic field, or viscous fluid. Later, in 1972,
Ruderman \cite{2} proposed that nuclear interactions within
extremely dense systems generate anisotropy. Since observations of
astrophysical structures reveal high nuclear density at their cores,
therefore, anisotropy is one of the salient features of their
intrinsic geometries and evolution. Researchers have considered
radial and tangential components of pressure to incorporate
anisotropy in the structure of stellar objects. Herrera and Santos
\cite{7} investigated possible factors that induce anisotropy in
spherical systems in GR. Static solutions describing the anisotropic
interior of cosmic objects were derived by Harko and Mak \cite{8}
through a particular form of anisotropy factor. Paul and Deb
\cite{9'} evaluated physically acceptable anisotropic solutions of
systems in hydrostatic equilibrium by considering observed masses of
compact stars. Murad \cite{9} incorporated the effects of the
electromagnetic field to model anisotropic strange stars by
considering a specific form of the metric potential.

Over the years, researchers have devised new techniques to obtain
viable models of stellar structures. Recently, Ovalle \cite{10}
proposed the method of minimal geometric deformation (MGD) to extend
a seed source (vacuum or isotropic) to complex fluid distributions.
This technique was first implemented in the framework of
Randall-Sundrum braneworld to derive consistent spherically
symmetric solutions. In this approach, an additional source is
incorporated in the seed distribution on the condition that the two
sources interact gravitationally only. A geometric deformation in
the radial metric component decouples the system of field equations
into two sets with lesser degrees of freedom as compared to the
original system. The two systems are solved independently and their
respective solutions are combined to obtain a solution of the
complete model.

Following the procedure of MGD, Ovalle and Linares \cite{11}
computed the braneworld version of Tolman IV and inspected the bulk
effects on the compactness of self-gravitating objects. Ovalle et
al. \cite{13} employed this technique to incorporate the effects of
anisotropy in perfect fluid configuration and generated three
anisotropic models from the Tolman IV solution. Gabbanelli et al.
\cite{15} discussed the salient features of anisotropic version of
the Durgapal-Fuloria solution. Estrada and Tello-Ortiz \cite{16}
adopted the MGD approach to construct two physically acceptable
anisotropic solutions from Heintzmann solution. Sharif and Sadiq
\cite{17} applied this method to Krori-Barua (KB) solution and
explored the impact of charge on the extended anisotropic system.
Geometric deformations on Tolman VII metric potentials have also
been applied to construct a physically viable anisotropic solution
\cite{16a}. Sharif and Ama-Tul-Mughani \cite{17'} decoupled the
field equations representing a cloud of strings and obtained
corresponding anisotropic extensions.

Although gravitational decoupling via MGD is a highly effective
scheme for constructing viable solutions of the field equations.
However, deformation in the radial metric component splits the field
equations only when the exchange of energy and momentum between the
considered sources is restricted. In order to overcome this
shortcoming, Casadio et al. \cite{40} modified the MGD technique by
introducing deformations in radial as well as temporal metric
components. However, this extension is valid only in the absence of
matter and does not satisfy the Bianchi identities corresponding to
self-gravitating systems filled with fluid. Recently, Ovalle
\cite{41} presented the most general way of decoupling a spherical
system by geometrically deforming both (temporal/radial) metric
functions. The main advantage of extended geometric deformation
(EGD) decoupling is that it works for all regions of spacetime
without imposing any restriction on the choice of matter
distribution. Contreras and Bargue\~{n}o \cite{42} employed this
technique in 2+1-dimensions and extended vacuum BTZ solution to an
exterior charged BTZ solution. Sharif and Ama-Tul-Mughani
implemented EGD approach to generate anisotropic analogues of Tolman
IV \cite{43} and KB \cite{44} solutions. Recently, MGD as well as
EGD approaches have been used to obtain anisotropic solutions in
modified theories as well \cite{28}.

In 1937, Dirac \cite{18'} hypothesized that all large numbers
obtained by the combinations of fundamental atomic constants are
related to cosmological parameters. Subsequently, the gravitational
constant $G$ must be a function of cosmic time. In 1961, Brans and
Dicke \cite{18} modified GR by incorporating Dirac observations in a
scalar-tensor theory and formulated a spherical vacuum solution.
Brans-Dicke (BD) gravity incorporates a massless scalar field
$\varphi=\frac{1}{G(t)}$ to discuss the evolution of the cosmos. A
tunable parameter ($\omega_{BD}$) couples the scalar field to the
matter distribution. As the role of scalar field is enhanced during
the inflationary era, the values of the coupling parameter must be
small to explain this scenario \cite{19}. On the other hand, the
solar system tests are satisfied for $\omega_{BD}>40,000$ \cite{20}.
This issue is resolved by the self-interacting BD (SBD) theory which
assigns a mass to the scalar field through a potential function
$V(\Psi)$ (where $\Psi$ is a massive scalar field) \cite{21}. In SBD
theory, if the mass of the scalar field is greater than
$2\times10^{-25}GeV$, the solar system observations cannot constrain
$\omega_{BD}$ and its values greater than $-\frac{3}{2}$ are allowed
\cite{23'}.

Solutions representing different scenarios have been formulated in
BD theory. Buchdahl \cite{22'} considered spherical as well as
axially symmetric spacetimes to show that a static vacuum solution
of GR can generate a family of static vacuum solutions in BD theory.
Sneddon and McIntosh \cite{22} extended Buchdahl work by applying
Geroch method \cite{23} to construct new vacuum solutions. Bruckman
and Kazes \cite{24} applied a linear equation of state (EoS) to
perfect fluid model and formulated an exact spherical solution with
infinite density at the center. Goswami \cite{25} constructed a
class of vacuum solutions by converting the BD field equations to
Einstein vacuum field equations. This work was extended to evaluate
solutions in the presence of electromagnetic field  as well as an
irrotational barotropic fluid \cite{25'}. A Demia\'{n}ski-type
metric was obtained via a complex coordinate transformation by Krori
and Bhattacharjee \cite{25a}. Riazi and Askari \cite{26}
approximated spherically symmetric solutions for a static vacuum
spacetime and examined the behavior of rotation curves. Recently,
isotropic versions of Durgapal-Fuloria and KB solutions were
extended to anisotropic domain through MGD approach in the context
of SBD theory \cite{27}.

The realistic models of relativistic stars have extensively been
discussed in scalar-tensor theories. Yazadjiev et al. \cite{55}
explored how the structure of slowly rotating neutron stars deviate
from the GR model in the presence of a massive scalar field.
Ramazano\~{g}lu and Pretorius \cite{56} reviewed the range allowed
for mass and scalarization of neutron stars by allotting a mass to
the scalar field. Doneva and Yazadjiev \cite{57} investigated the
dynamics of rapidly rotating neutron stars in the presence of a
massive scalar field and concluded that deviations from GR can be
large due to moment of inertia. Staykov et al. \cite{58} extended
this work by considering a self-interacting potential along with a
massive scalar field to analyze the behavior of static and slowly
rotating neutron stars. Popchev et al. \cite{59} analyzed the
effects of a self-interacting massive scalar field on moment of
inertia and compactness of slowly rotating neuron stars by employing
different EoS.

The focus of this paper is to evaluate viable anisotropic versions
of Tolman IV and KB solutions by decoupling the SBD field equations
via the EGD approach. The paper is organized as follows. In section
\textbf{2}, we construct the field equations by introducing an
anisotropic source in perfect fluid distribution which are then
decoupled in section \textbf{3}. Extended anisotropic solutions are
computed through some physical constraints and examined for
viability in section \textbf{4}. In the last section, the main
results are summarized.

\section{Self-interacting Brans-Dicke Theory}

The modified action of SBD theory with an additional source in
relativistic units ($8\pi G_0=1$) is given by
\begin{equation}\label{0}
S=\int\sqrt{-g}(\mathcal{R}\Psi-\frac{\omega_{BD}}{\Psi}\nabla^{\gamma}\nabla_{\gamma}\Psi
-V(\Psi)+\emph{L}_m+\alpha\emph{L}_\Theta)d^{4}x,
\end{equation}
where $\mathcal{R}$ is the Ricci scalar, $\emph{L}_m$ and
$\emph{L}_\Theta$ represent Lagrangian densities of matter and new
source, respectively. Moreover, the additional source
($\Theta_{\gamma\delta}$) is coupled to the matter distribution
through a dimensionless parameter $\alpha$. The new source generally
induces anisotropy in an isotropic self-gravitating system by
including scalar, vector or tensor fields in the stellar model. The
SBD field equations and wave equation, obtained from the above
action, are respectively given as
\begin{eqnarray}\label{1}
G_{\gamma\delta}&=&T^{\text{(eff)}}_{\gamma\delta}+\frac{\alpha}{\Psi}\Theta_{\gamma\delta}
=\frac{1}{\Psi}(T_{\gamma\delta}^{(m)}
+T_{\gamma\delta}^\Psi+\alpha\Theta_{\gamma\delta}),\\\label{2}
\Box\Psi&=&\frac{\bar{T}}{3+2\omega_{BD}}+\frac{1}{3+2\omega_{BD}}
(\Psi\frac{dV(\Psi)}{d\Psi}-2V(\Psi)),
\end{eqnarray}
where $\Box$ denotes the d'Alembertian operator. The interior
configuration of a compact object filled with perfect fluid is
represented by the following energy-momentum tensor
\begin{equation}\label{3}
T_{\gamma\delta}^{(m)}=(\rho+p)u_{\gamma}u_{\delta}-pg_{\gamma\delta},
\end{equation}
where $\rho$, $p$ and $u_{\gamma}$ denote energy density, isotropic
pressure and four velocity, respectively. Also,
$\bar{T}=\Theta+T^{(m)}$,
($\Theta=g^{\gamma\delta}\Theta_{\gamma\delta},~T^{(m)}=g^{\gamma\delta}T_{\gamma\delta}^{(m)}$).
The energy-momentum tensor related to the massive scalar field is
defined as
\begin{equation}\label{4}
T_{\gamma\delta}^\Psi=\Psi_{,\gamma;\delta}-g_{\gamma\delta}\Box\Psi+\frac{\omega_{BD}}{\Psi}
(\Psi_{,\gamma}\Psi_{,\delta}
-\frac{g_{\gamma\delta}\Psi_{,\alpha}\Psi^{,\alpha}}{2})-\frac{V(\Psi)g_{\gamma\delta}}{2}.
\end{equation}

The internal geometry of a static spherical object is described by
the line element
\begin{equation}\label{5}
ds^2=e^{\upsilon(r)}dt^2-e^{\chi(r)}dr^2-r^2(d\theta^2+\sin^2\theta
d\phi^2).
\end{equation}
The field equations incorporating the anisotropic source are
formulated via Eqs.(\ref{1})-(\ref{5}) as
\begin{eqnarray}\label{6}
\frac{1}{r^2}-e^{-\chi}\left(\frac{1}{r^2}-\frac{\chi'}{r}\right)&=&
\frac{1}{\Psi}(\rho+\alpha\Theta_0^0+T_0^{0\Psi}),\\\label{7}
-\frac{1}{r^2}+e^{-\chi}\left(\frac{1}{r^2}+\frac{\upsilon'}{r}\right)&=&
\frac{1}{\Psi}(p-\alpha\Theta_1^1-T_1^{1\Psi}),\\\label{8}
\frac{e^{-\chi}}{4}\left(2\upsilon''+\upsilon'^2-\chi'\upsilon'+2\frac{\upsilon'-\chi'}{r}\right)&=&
\frac{1}{\Psi}(p-\alpha\Theta_2^2-T_2^{2\Psi}),
\end{eqnarray}
where
\begin{eqnarray}\nonumber
T_0^{0\Psi}&=&e^{-\chi}\left[\Psi''+\left(\frac{2}{r}-\frac{\chi'}{2}
\right)\Psi'+\frac{\omega_{BD}}{2\Psi}\Psi'^2-e^\chi\frac{V(\Psi)}
{2}\right],\\\nonumber
T_1^{1\Psi}&=&e^{-\chi}\left[\left(\frac{2}{r}+\frac{\upsilon'}
{2}\right)\Psi'-\frac{\omega_{BD}}{2\Psi}\Psi'^2-e^\chi\frac{V(\Psi)}{2})\right],\\\nonumber
T_2^{2\Psi}&=&e^{-\chi}\left[\Psi''+\left(\frac{1}{r}-\frac{\chi'}
{2}+\frac{\upsilon'}{2}\right)\Psi'+\frac{\omega_{BD}}{2\Psi}\Psi'^2-e^\chi\frac{V(\Psi)}{2}
\right].
\end{eqnarray}
Here prime denotes differentiation with respect to $r$. The
evolution equation (\ref{2}) for the metric (\ref{5}) turns out to
be
\begin{eqnarray}\label{2*}
\Box\Psi&=&-e^{-\chi}\left[\left(\frac{2}{r}-\frac{\chi'} {2}
+\frac{\upsilon'}{2}\right)\Psi'+\Psi''\right].
\end{eqnarray}
For $\Theta_1^1\neq\Theta_2^2$, the anisotropy introduced by the
extra source is $\Delta=T_2^{2\text{(eff)}}-T_1^{1\text{(eff)}}$.
This work is carried out by choosing the following potential
function
\begin{equation*}
V(\Psi)=\frac{1}{2}m_{\Psi}^2\Psi^2,
\end{equation*}
where $m_{\Psi}$ is the mass of the scalar field.

\section{Gravitational Decoupling}

Equations (\ref{6})-(\ref{2*}) form a system of non-linear
differential equations with eight unknowns: two metric potentials
($\upsilon,~\chi$), five matter variables ($\rho,~p,~\Theta_{0}^{0},\\
\Theta_{1}^{1},~\Theta_{2}^{2}$) and a massive scalar field. In
order to evaluate the unknown functions, we implement the novel
technique of EGD \cite{41} on SBD field equations. This technique
determines the effect of $\Theta_\delta^\gamma$ on the matter
distribution by inducing the following deformations in the metric
potentials
\begin{eqnarray}\label{11}
\upsilon(r)&\mapsto&\mu(r)+\alpha g(r),\\\label{11'}
e^{-\chi(r)}&\mapsto&e^{-\eta(r)}+\alpha f(r),
 \end{eqnarray}
where $f(r)$ and $g(r)$ encode the translations in radial and
temporal metric components, respectively. Moreover, the free
parameter $\alpha$ controls the contribution of deformations. It is
noteworthy that spherical symmetry of the compact object is
preserved under these geometric deformations. Substituting the
deformed metric potentials in Eqs.(\ref{6})-(\ref{8}) splits the
system into two sets. The first set corresponds to $\alpha=0$ and
exclusively describes the isotropic configuration as
\begin{eqnarray}\nonumber
\rho&=&\frac{1}{2r^2\Psi(r)}\left\{e^{-\eta(r)}\left(r^2e^{\eta(r)}V(\Psi)\Psi(r)+r^2
(-\omega_{BD})\Psi'^2(r)+\left(\left(r\eta'(r)\right.\right.\right.\right.\\\label{12}
&-&\left.\left.\left.\left.4\right)\Psi'(r)
-2r\Psi''(r)\right)r\Psi(r)+2\Psi^2(r)\left(r\eta'(r)+e^{\eta(r)}-1\right)\right)\right\}
,\\\nonumber
p&=&\frac{1}{2}\left\{\frac{1}{r^2\Psi(r)}\left(e^{-\eta(r)}\left(-r^2\omega_{BD}
\Psi'^2(r)+\Psi^2(r)\left(2r\mu'(r)-2e^{\eta(r)}+2\right)\right.\right.\right.\\\label{13}
&+&\left.\left.\left.r\Psi(r)
\left(r\mu'(r)+4\right)\Psi'(r)\right)\right)-V(\Psi)\right\},\\\nonumber
p&=&\frac{1}{4r\Psi(r)}\left\{e^{-\eta(r)}\left(2\Psi(r)\left(\Psi'(r)\left(r\mu'(r)-r
\eta'(r)+2\right)+2r\Psi''(r)\right)+\Psi^2(r)\right.\right.\\\nonumber
&\times&\left.\left.\left(2r\mu''(r)+\mu'(r)
\left(2-r\eta'(r)\right)+r\mu'^2(r)-2\eta'(r)\right)-2re^{\eta(r)}\Psi(r)
V(\Psi)\right.\right.\\\label{14}
&+&\left.\left.2r\omega_{BD}\Psi'^2(r)\right)\right\}.
\end{eqnarray}
The conservation of isotropic matter distribution in ($\mu,\eta$)
coordinates is represented by the conservation equation
\begin{equation}\label{14*}
T^{1'(\text{eff})}_{1}-\frac{\mu'(r)}{2}
(T^{0(\text{eff})}_{0}-T^{1(\text{eff})}_{1})=0.
\end{equation}

The second set containing evolution equations for the anisotropic
source is given as
\begin{eqnarray}\nonumber
\Theta_0^0&=&\frac{-1}{2r^2\Psi (r)}\left\{\left
(r\Psi(r)f'(r)\left(r\Psi'(r)+2\Psi
(r)\right)+f(r)\left(r^2\omega_{BD}\Psi'^2(r)\right.\right.\right.\\\label{15}
&+&\left.\left.\left.2r\Psi(r)\left(r\Psi
''(r)+2\Psi'(r)\right)+2\Psi^2(r)\right)\right)\right\},\\\nonumber
\Theta_1^1&=&\frac{-
f(r)}{2r^2\Psi(r)}\left(-r^2\omega_{BD}\Psi'^2(r)+r
\Psi(r)\left(r\upsilon'(r)+4\right)\Psi'(r)+2\Psi^2(r)\left(r\upsilon'(r)\right.\right.\\\label{16}
&+&\left.\left.1\right)\right) -\frac{
e^{-\eta(r)}g'(r)\left(r\Psi'(r)+2\Psi(r)\right)}{2r},\\\nonumber
\Theta_2^2&=&\frac{-
f(r)}{4r\Psi(r)}\left(2\Psi(r)\left(\left(r\upsilon'(r)+2\right)\Psi'(r)+2
r\Psi''(r)\right)+\Psi^2(r)\left(2r\upsilon''(r)\right.\right.\\\nonumber
&+&\left.\left.r\upsilon'^2(r)+2\upsilon'(r)\right)+2
r\omega_{BD}\Psi'^2(r)\right)-\frac{
f'(r)}{4r}\left(\Psi(r)\left(r\upsilon'(r)+2\right)\right.\\\nonumber
&+&\left.2r\Psi'(r)\right)-\frac{
e^{-\eta(r)}}{4r}\left(2rg'(r)\Psi'(r)+\Psi(r)\left(2rg''(r)+\alpha
rg'^2(r)+g'(r)\right.\right.\\\label{17}
&\times&\left.\left.\left(2r\mu'(r)-r\eta'(r)+2\right)\right)\right).
\end{eqnarray}
The divergence of the source $\Theta^{\gamma}_{\delta}$ leads to
\begin{equation}\label{17*}
\Theta^{1'(\text{eff})}_{1}
-\frac{\upsilon'(r)}{2}(\Theta^{0(\text{eff})}_{0}-\Theta^{1(\text{eff})}_{1})
-\frac{2}{r}(\Theta^{2(\text{eff})}_{2}-\Theta^{1(\text{eff})}_{1})
=\frac{g'(r)}{2}(T^{0(\text{eff})}_{0}-T^{1(\text{eff})}_{1}),
\end{equation}
where
\begin{eqnarray*}
\Theta^{0(\text{eff})}_0&=&\frac{1}{\Psi}\left(\Theta^0_0+\frac{1}{2}
f'(r)\Psi'(r)+f(r)\Psi''+\frac{\omega_{BD}f(r)\Psi'^2}{2\Psi}
+\frac{2f(r)\Psi '(r)}{r}\right),\\
\Theta^{1(\text{eff})}_1&=&\frac{1}{\Psi}\left(\Theta^1_1+\frac{1}{2r\Psi}
e^{-\eta(r)}\Psi'(r)\left(f(r)e^{\eta(r)}\left(\Psi(r)\left(r\upsilon'(r)+4\right)-r
\omega_{BD}\right.\right.\right.\\
&\times&\left.\left.\left.\Psi'(r)\right)+r\Psi(r)g'(r)\right)\right),\\
\Theta^{2(\text{eff})}_2&=&\frac{1}{\Psi}\left(\Theta^2_2+\frac{1}{2r\Psi}
e^{-\eta(r)}\left(r\Psi(r)\Psi'(r)\left(e^{\eta(r)}f'(r)+g'(r)\right)+f(r)\right.\right.\\&\times&\left.\left.
e^{\eta(r)}\left(\Psi(r)\left(\left(r\upsilon'(r)+2\right)\Psi'(r)+2r\Psi''(r)\right)+r
\omega_{BD}\Psi'^2(r)\right)\right)\right).
\end{eqnarray*}
The conservation equation of the energy-momentum tensor
$T^{\gamma(\text{eff})}_{\delta}$ in ($\upsilon,\chi$)-coordinate
system yields
\begin{equation}\label{18*}
\nabla_{\gamma}T^{\gamma(\text{eff})}_{\beta}=\nabla_{\gamma}^{(\mu,\eta)}T^{\gamma(\text{eff})}_{\beta}
-\frac{g'(r)}{2}(T^{0(\text{eff})}_{0}-T^{1(\text{eff})}_{1})\delta^1_{\beta},
\end{equation}
where $\nabla_{\gamma}^{(\mu,\eta)}$ represents the divergence of a
tensor in ($\mu,\eta$)-frame. As a direct consequence of
Eqs.(\ref{14*}) and (\ref{17*}), we have
\begin{equation}\label{19}
\nabla_{\gamma}^{(\mu,\eta)}T^{\gamma(\text{eff})}_{\beta}=0,\quad
\nabla_{\gamma}\Theta^{\gamma(\text{eff})}_{\beta}
=\frac{g'(r)}{2}(T^{0(\text{eff})}_{0}-T^{1(\text{eff})}_{1})\delta^1_{\beta}.
\end{equation}

Equations (\ref{18*}) and (\ref{19}) imply that exchange of energy
takes place between the sources $T^{(m)}_{\gamma\delta}$ and
$\Theta_{\gamma\delta}$ but the overall energy and momentum of the
system remain unchanged. Thus, these sources can be decoupled
provided that energy can be transferred from one setup to the other.
However, if $T^{(m)}_{\gamma\delta}$ represents either a vacuum
solution or a barotropic fluid, matter sources interacting only
gravitationally can also be decoupled via EGD approach. It is
worthwhile to mention here that in the specific case of MGD
$(g(r)=0)$, there is no exchange of matter between the isotropic and
anisotropic configurations.

\section{Anisotropic Solutions}

When we apply the EGD technique, the system (\ref{6})-(\ref{8}) is
decomposed into two sets: Eqs.(\ref{12})-(\ref{14}) represent the
seed source in terms of $T^{(m)}_{\gamma\delta},~\mu$ and $\eta$
whereas the influence of the additional source is determined by the
second set (\ref{15})-(\ref{17}) with five unknowns
$(g(r),~f(r),~\Theta^0_0,~\Theta^1_1,~\Theta^2_2)$. The undetermined
variables of the second set can be evaluated if a viable solution
for the isotropic sector is known. Thus, EGD approach has simplified
the process of extracting solutions of the field equations by
reducing the degrees of freedom from 4 to 2. In this section, we
obtain anisotropic analogues of two solutions: Tolman IV and KB.

In 1939, Tolman \cite{30} constructed eight static spherically
symmetric solutions for perfect fluid and explored the conditions
for smooth matching of interior and exterior geometries. Tolman IV
is one of the physically acceptable solutions \cite{29c} which
corresponds to a non-vanishing surface density. It has previously
been employed to investigate different features of self-gravitating
systems \cite{43, 30a}. The line element of Tolman IV solution is
written as
\begin{eqnarray}\label{18}
ds^2=B^2(1+\frac{r^2}{A^2})dt^2-\frac{1+\frac{2r^2}{A^2}}{(1+\frac{r^2}{A^2})
(1-\frac{r^2}{F^2})}dr^2-r^2(d\theta^2+\sin^2\theta d\phi^2),
\end{eqnarray}
where the constants $A,~B$ and $F$ are determined through the
matching of internal and external spacetimes at the boundary
($\Sigma$) of the celestial object. The Schwarzschild metric
describes vacuum in the exterior of the celestial object as
\begin{equation}\label{20*}
ds^2=(1-\frac{2M}{r})dt^2-\frac{1}{(1-\frac{2M}{r})}dr^2
-r^2(d\theta^2+\sin^2\theta d\phi^2),
\end{equation}
where $M$ represents the overall mass of the compact structure. The
junction conditions that ensure smooth matching of internal and
external geometries at the stellar boundary ($r=R=$ radius of
compact object) are expressed as
\begin{eqnarray*}
(g_{\gamma\delta}^-)_{\Sigma}&=&(g_{\gamma\delta}^+)_{\Sigma},\quad
(p_{r})_{\Sigma}=0,\\
(\Psi^-(r))_\Sigma&=&(\Psi^+(r))_\Sigma,\quad
(\Psi'^-(r))_\Sigma=(\Psi'^+(r))_\Sigma.
\end{eqnarray*}
The junction conditions evaluate the constants $A,~B$ and $F$ (for
$\alpha=0$) as
\begin{eqnarray}\label{20}
A^2&=&-\frac{R^2\left(R^2\zeta+M\left(28R-2
R\zeta\right)+2M^2(\omega_{BD}-12)-8
R^2\right)}{R\zeta(R-2M)+2M^2\omega_{BD}},\\\label{21}
B^2&=&\frac{(R-2M)\left(3R^2\zeta-6M\left(R
\zeta+10R\right)+6M^2(\omega_{BD}+12)+8
R^2\right)}{2R\left(R^2\zeta-2M\left(R
\zeta+11R\right)+2M^2(\omega_{BD}+12)+4R^2\right)},\\\nonumber
F^2&=&(4R^3(3M-2R)\left(\zeta+4\right))(m_\Psi^4R^6+4R^2\zeta
+M^2\left(4m_\Psi^4R^4+2\zeta(\omega_{BD}\right.\\\label{22}
&+&\left.12)+8(\omega_{BD}+12)\right)-4M\left(m_\Psi^4R^5+6R
\zeta+16R\right))^{-1}.
\end{eqnarray}
where $\zeta=m_\Psi^2R^2\sqrt{1-\frac{2M}{R}}$.

Krori and Barua \cite{30b} formulated a physically acceptable
solution for a static charged sphere. The highlight of this solution
is that no restrictions are imposed on the metric functions to avoid
singularities, i.e., it is regular throughout the spacetime. This
solution has proved helpful in checking the impact of
electromagnetic field on matter source. However, researchers have
also employed this ansatz to inspect physical characteristics of
uncharged systems \cite{31'}. The KB solution is defined by the
following line element
\begin{eqnarray}\label{18'}
ds^2&=&e^{ar^2+b}dt^2-e^{cr^2}dr^2-r^2(d\theta^2+\sin^2\theta
d\phi^2),
\end{eqnarray}
where the constants $a,~b$ and $c$ are evaluated (for $\alpha=0$)
through the matching conditions as
\begin{eqnarray}\label{24}
a&=&\frac{R\zeta(R-2 M)+2M^2\omega_{BD}}{4R^2
(R-2M)(2R-3M)},\\\label{25}
b&=&\frac{R\zeta (R-2M)+2
M^2\omega_{BD}}{-4\left(6M^2-7M
R+2R^2\right)}+\ln\left(1-\frac{2M}{R}\right),\\\label{26}
c&=&-\frac{\ln\left(1-\frac{2M}{R}\right)}{R^2}.
\end{eqnarray}
The anisotropic model is completely specified by the following
matter variables
\begin{eqnarray}\nonumber
\rho&=&\frac{e^{-\eta(r)}}{2r^2\Psi
(r)}\left(-r\Psi(r)\left(\Psi'(r)\left(\alpha r
e^{\eta(r)}f'(r)+4\alpha
f(r)e^{\eta(r)}-r\eta'(r)+4\right)\right.\right.\\\nonumber&+&\left.\left.2r\Psi
''(r)\left(\alpha f(r)e^{\eta(r)}+1\right)\right)-2\Psi^2(r)
\left(\alpha re^{\eta(r)}f'(r)+\alpha
f(r)e^{\eta(r)}\right.\right.\\\nonumber&-&\left.\left.r\eta'(r)-e^{\eta
(r)}+1\right)+r^2(-\omega_{BD})\Psi'^2(r)\left(\alpha f(r)e^{\eta
(r)}+1\right)\right.\\\label{28}&+&\left.r^2e^{\eta(r)}\Psi(r)V(\Psi)\right),\\\nonumber
p_r&=&\frac{\Psi(r)}{r^2}\left(\left(\alpha f(r)+e^{-\eta
(r)}\right)\left(\alpha
rg'(r)+r\mu'(r)+1\right)-1\right)-\frac{1}{2r\Psi
(r)}\\\nonumber&\times&\left(\Psi'(r)\left(\alpha
f(r)+e^{-\eta(r)}\right)\left(r\omega_{BD}\Psi'(r)-\Psi(r)
\left(\alpha
rg'(r)+r\mu'(r)+4\right)\right)\right)\\\label{29}&-&\frac{V(\Psi)}{2},\\\nonumber
p_\perp&=&\left(\alpha
f(r)+e^{-\eta(r)}\right)\left(\frac{1}{2}\Psi'(r)\left(\frac{\alpha
e^{\eta(r)}f'(r)-\eta'(r)}{\alpha f(r)e^{\eta(r)}+1}+\alpha
g'(r)+\mu'(r)\right.\right.\\\nonumber&+&\left.\left.\frac{2}{r}\right)+\Psi''(r)+\frac{\omega_{BD}\Psi'^2(r)}{2
\Psi(r)}\right)+\frac{1}{2}\Psi(r)\left(\alpha f(r)+e^{-\eta
(r)}\right)\left((\left(\alpha
e^{\eta(r)}f'(r)\right.\right.\\\nonumber&-&\left.\left.\eta'(r)\right)
\left(\alpha g'(r)+\mu'(r)\right))(2\alpha
f(r)e^{\eta(r)}+2)^{-1}+\frac{1}{r}(\frac{\alpha
e^{\eta(r)}f'(r)-\eta'(r)}{\alpha
f(r)e^{\eta(r)}+1}\right.\\\label{30}&+&\left.\alpha
g'(r)+\mu'(r))+\alpha g''+\frac{1}{2} \left(\alpha
g'(r)+\mu'^2(r)\right)+\mu''(r)\right)-\frac{V(\Psi)}{2},
\end{eqnarray}
with anisotropy $\Delta=p_\perp-p_r$.

In order to extend the seed solutions to the anisotropic domain, we
require two constraints on $\Theta^{\gamma}_{\delta}$ to close the
anisotropic system. For this purpose, we choose a mimic constraint
\begin{equation}\label{27}
\Theta^1_1(r)=p(r),
\end{equation}
which fulfills the requirement of vanishing pressure at the
hypersurface. Under this constraint, the values of the constants $F$
and $a$ remain unchanged. The remaining constants $A$ and $c$ appear
as free parameters in corresponding extended versions whose values
are chosen as presented in Eqs.(\ref{20}) and (\ref{24}),
respectively. For the second constraint, a linear EoS as well as a
regularity condition on anisotropy is implemented which will be
discussed in subsections \textbf{4.1} and \textbf{4.2},
respectively.

The limits enforced by the weak-field on values of the coupling
parameter can be avoided through a lower bound for mass of the
scalar field ($m_\Psi>10^{-4}$ in dimensionless units). In
accordance with this limit, we take $m_{\Psi}=0.01$ and solve the
wave equation numerically to determine the massive scalar field.
Different features of anisotropic models are investigated
graphically for three values of $\alpha$ (0.2, 0.55, 0.9) by
employing the observed mass ($1.97M_{\bigodot}$) and radius
($11.29$km) of the star PSR J1614-2230.

\subsection{Case I: Linear Equation of State}

We consider a linear EoS for the source $\Theta_{\delta}^{\gamma}$
as
\begin{equation}\label{29'}
\Theta_0^0=\chi\Theta^1_1+\psi\Theta^2_2.
\end{equation}
Setting $\chi=1$ and $\psi=0$ in the above equation leads to
\begin{eqnarray}\nonumber
&&\frac{e^{-\eta(r)}}{r\Psi(r)}\left(-r\Psi(r)\left(\Psi'(r)\left(re^{\eta(r)}
f'(r)+4f(r)e^{\eta(r)}+r\mu'(r)+4\right)+2rf(r)\right.\right.\\\nonumber
&&\times\left.\left.e^{\eta(r)}\Psi''(r)\right)-2\Psi^2(r)\left(re^{\eta(r)}f'(r)+f(r)e^{\eta(r)}+r\mu
'(r)-e^{\eta(r)}+1\right)\right.\\\label{30''}
&&+\left.r^2(-\omega_{BD})\left(f(r)e^{\eta(r)}-1\right)
\Psi'^2(r)+r^2e^{\eta(r)}\Psi(r)V(\Psi)\right)=0,
\end{eqnarray}
which is solved numerically for $f(r)$ along with the wave equation
with the central conditions $\Psi(0)=0.2,~\Psi'(0)=0$ and $f(0)=0$.
On the other hand, the mimic constraint (\ref{27}) yields the
following temporal geometric function
\begin{eqnarray}\nonumber
g(r)&=&\int\left((r^2\omega_{BD}\left(f(r)e^{\eta(r)}+1\right)\Psi
'(r)^2-r\Psi(r)\left(f(r)e^{\eta(r)}+1\right)\left(r\mu'(r)\right.\right.\\\nonumber
&+&\left.\left.4\right)
\Psi'(r)-2\Psi^2(r)\left(f(r)e^{\eta(r)}\left(r\mu'(r)+1\right)+r\mu
'(r)-e^{\eta(r)}+1\right)\right.\\\nonumber&+&\left.r^2e^{\eta(r)}\Psi(r)V(\Psi))(r\Psi(r)
\left(r\Psi'(r)+2\Psi(r)\right)\left(\alpha f(r)e^{\eta
(r)}+1\right))^{-1}\right)dr.\\\label{30'}
\end{eqnarray}
Substituting the metric functions and constants corresponding to
Tolman IV solution in Eqs.(\ref{28})-(\ref{30}), (\ref{30''}) and
(\ref{30'}) provides the extended version of this solution.

The graphical analysis of state determinants is provided in Figure
\textbf{1} with $\omega_{BD}=9.87$. A stellar model is well-behaved
if the state parameters are positive, finite and decrease
monotonically away from the center. Moreover, radial pressure must
vanish at the boundary of the star. It is observed from Figure
\textbf{1} that energy density as well as pressure components are
positive throughout and maximum at the center for $\alpha=0.2$ and
0.55. However, for $\alpha=0.9$, the transverse pressure increases
monotonically instead of decreasing. The anisotropy is zero at the
center and increases towards the surface indicating the presence of
an outward repulsive force. It is noted that higher values of
$\alpha$ increase the density and repulsive force in the interior of
the structure whereas the pressure components decrease.
\begin{figure}\center
\epsfig{file=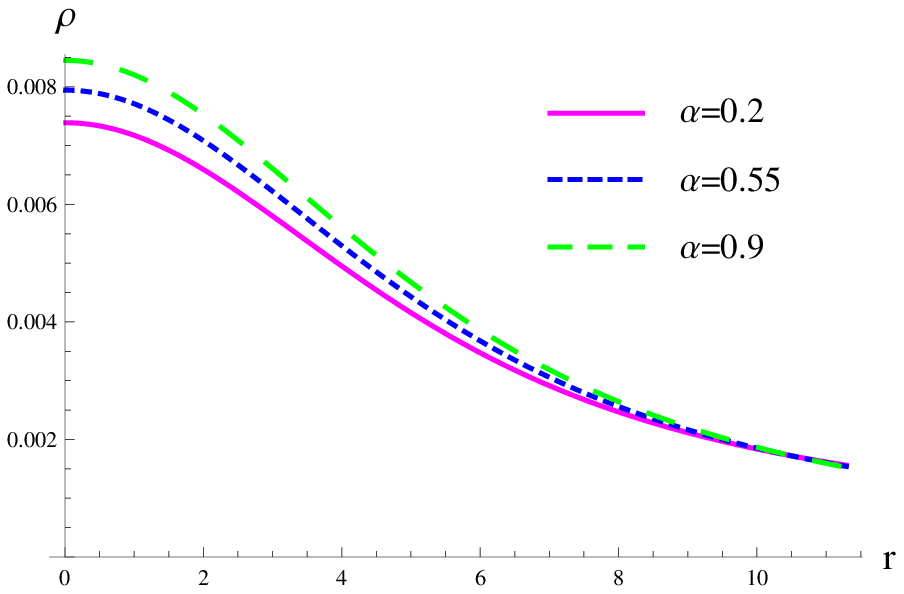,width=0.4\linewidth}\epsfig{file=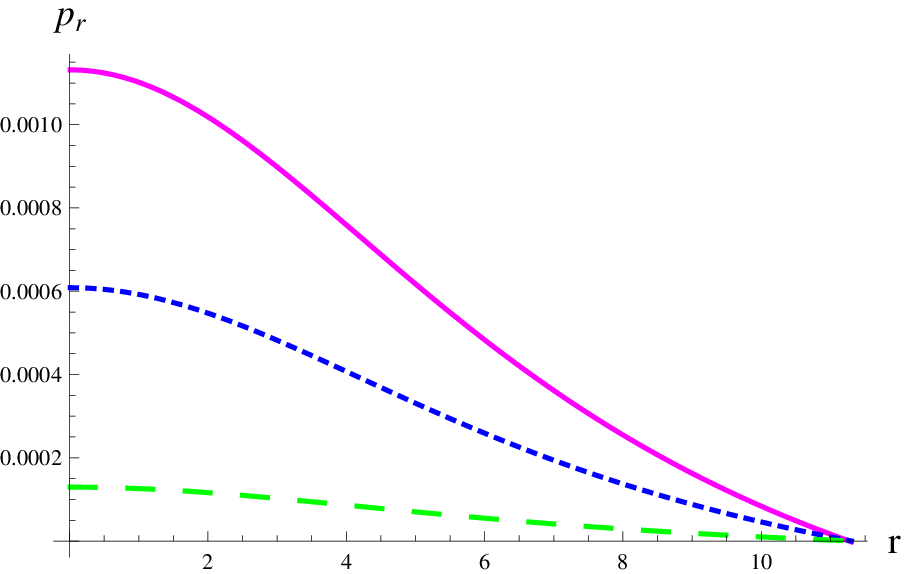,width=0.4\linewidth}
\epsfig{file=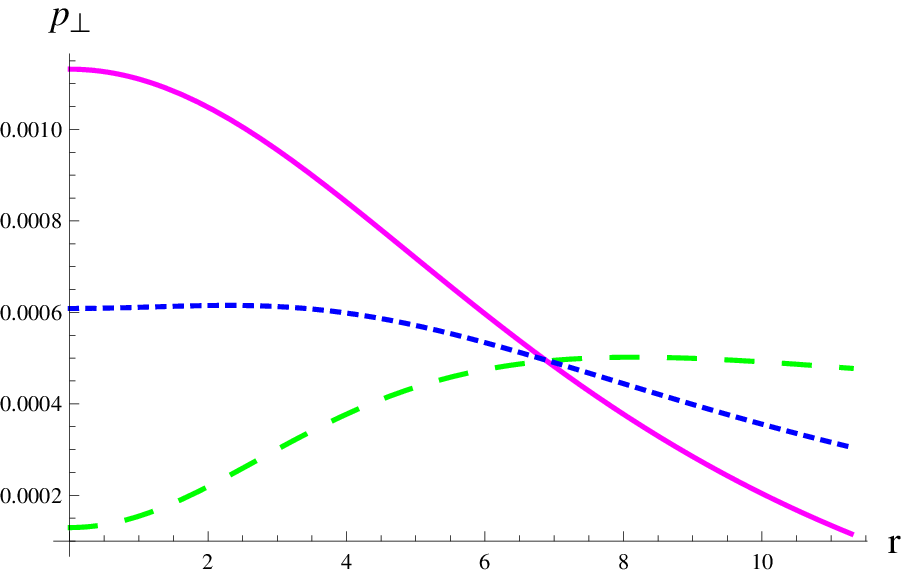,width=0.4\linewidth}\epsfig{file=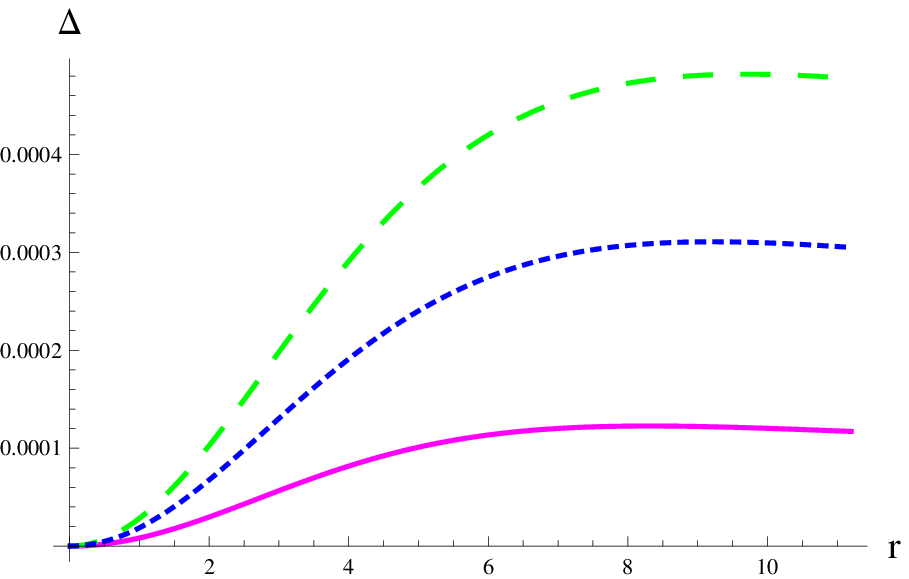,width=0.4\linewidth}
\caption{Plots of matter variables and anisotropy of extended Tolman
IV solution for case I.}
\end{figure}
\begin{figure}\center
\epsfig{file=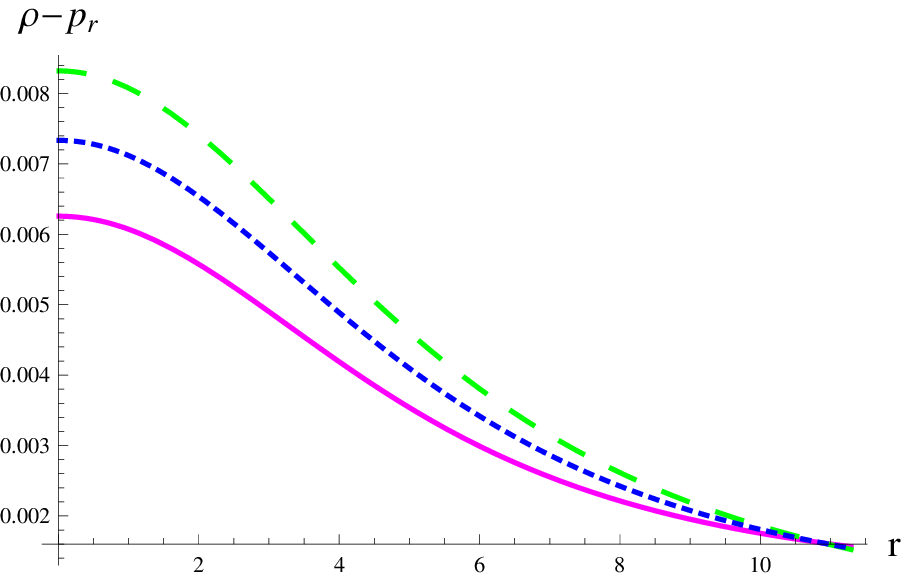,width=0.4\linewidth}\epsfig{file=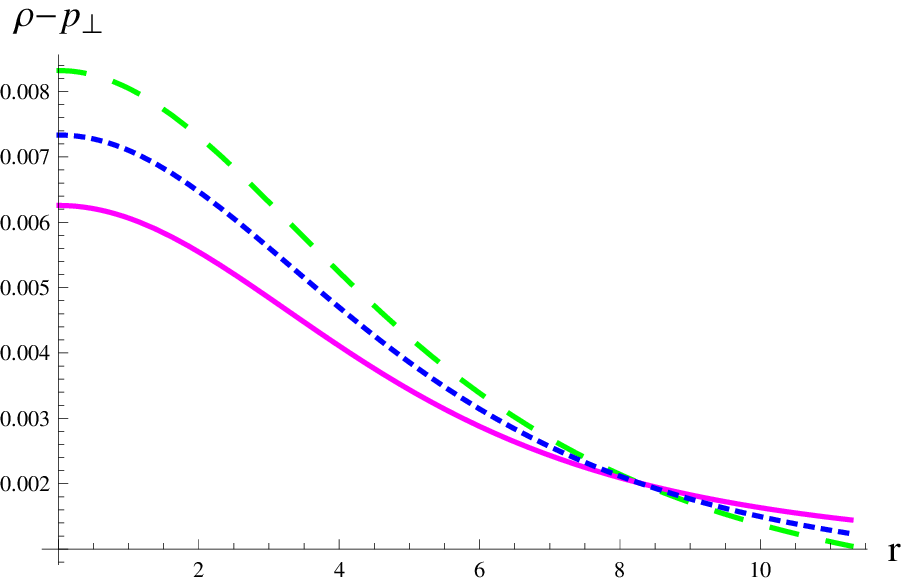,width=0.4\linewidth}
\caption{DEC for anisotropic Tolman IV with case I.}
\end{figure}

Four energy bounds on matter variables distinguish normal matter
from exotic material. Since stellar structures are composed of
ordinary matter, it is crucial that the parameters governing the
interior of compact objects obey these energy conditions. The null,
weak, strong and dominant energy conditions in the framework of SBD
theory are, respectively, expressed as \cite{34}
\begin{eqnarray*}
&&\text{NEC:}\quad\rho+p_r\geq0,\quad\rho+p_\perp\geq0,\\
&&\text{WEC:}\quad\rho\geq0,\quad\rho+p_r\geq0,\quad\rho+p_\perp\geq0,\\
&&\text{SEC:}\quad\rho+p_r+2p_\perp\geq0,\\
&&\text{DEC:}\quad\rho-p_r\geq0,\quad \rho-p_\perp\geq0.
\end{eqnarray*}
The first three conditions are readily satisfied for extended Tolman
IV solution as energy density and pressure (radial/transverse) are
positive within the compact object. Figure \textbf{2} demonstrates
that the parameters governing the matter source agree with DEC
ensuring viability of the model.

Another important physical feature of a self-gravitating system is
its compactness ($u(r)$) in a state of equilibrium. The compactness
factor is defined as the relation of mass to the radius of the
object. Buchdahl \cite{32} calculated the upper limit of this
parameter for a fluid with non-increasing energy density by matching
the interior of a static sphere to Schwarzschild exterior solution.
This limit is given as
\begin{equation*}
u(r)=\frac{m}{R}<\frac{4}{9},
\end{equation*}
where $ m(r)=\frac{R}{2}(1-e^{-\chi})$. The compactness factor
obtained for anisotropic Tolman IV solution (shown in Figure
\textbf{3}) conforms to Buchdahl limit. The surface redshift
($Z(r)$) of a celestial object gauges the increase in wavelength of
electromagnetic radiation due to gravitational force exerted by the
star. It is defined as
\begin{equation*}
Z(r)=\frac{1}{\sqrt{1-2u}}-1.
\end{equation*}
For a perfect fluid distribution, Buchdahl limit restricts the value
of redshift at the stellar surface as $Z(r)<2$. However, for an
anisotropic configuration, the upper limit of surface redshift
changes to 5.211 \cite{33}. It is observed from Figure \textbf{3}
that the range of redshift parameter complies with the above limit.
\begin{figure}\center
\epsfig{file=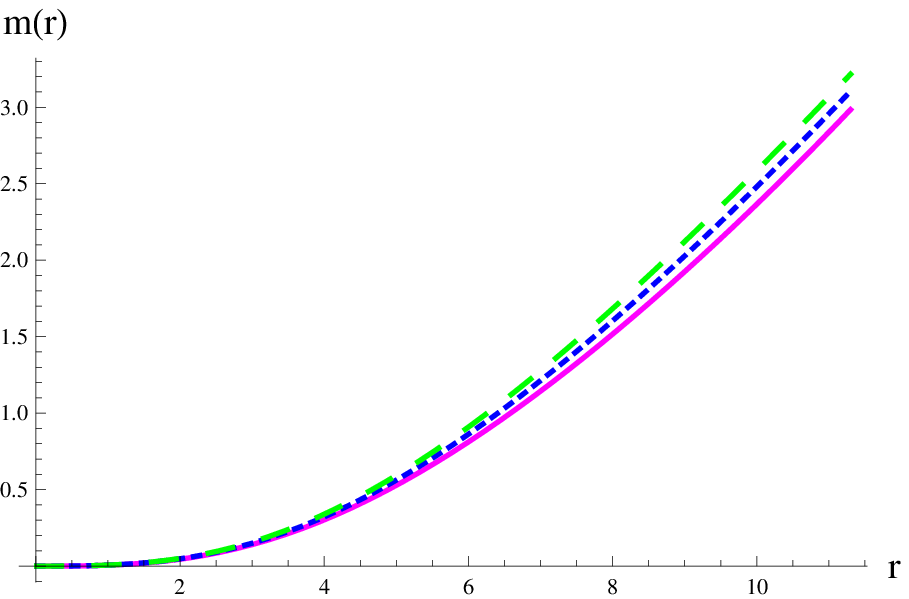,width=0.35\linewidth}\epsfig{file=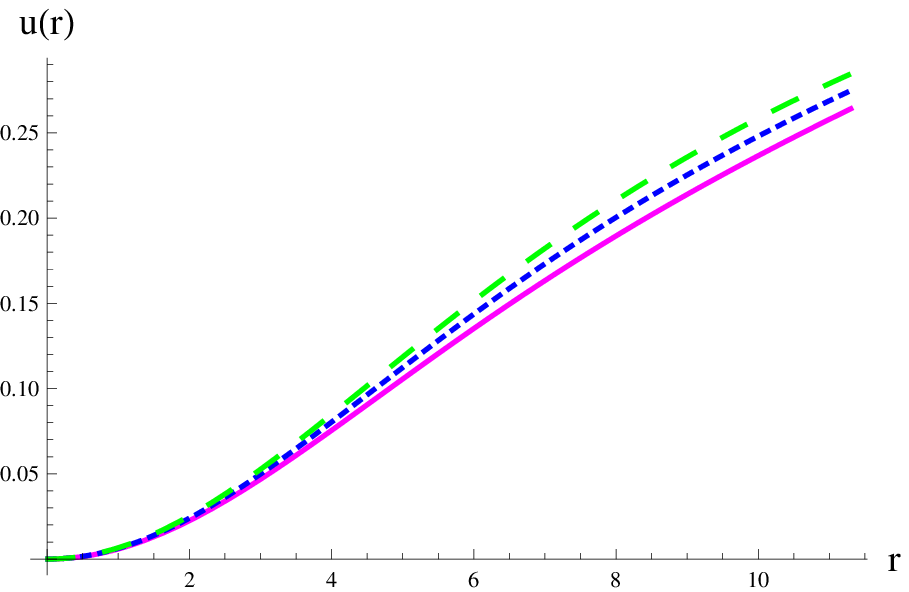,width=0.35\linewidth}
\epsfig{file=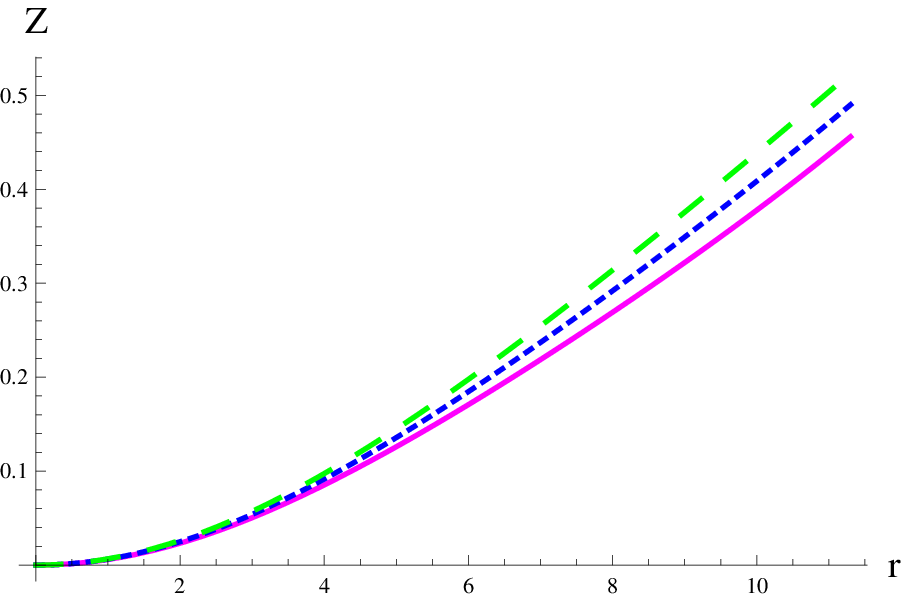,width=0.35\linewidth} \caption{Plots of mass,
compactness and redshift parameters corresponding to anisotropic
Tolman IV for case I.}
\end{figure}
\begin{figure}\center
\epsfig{file=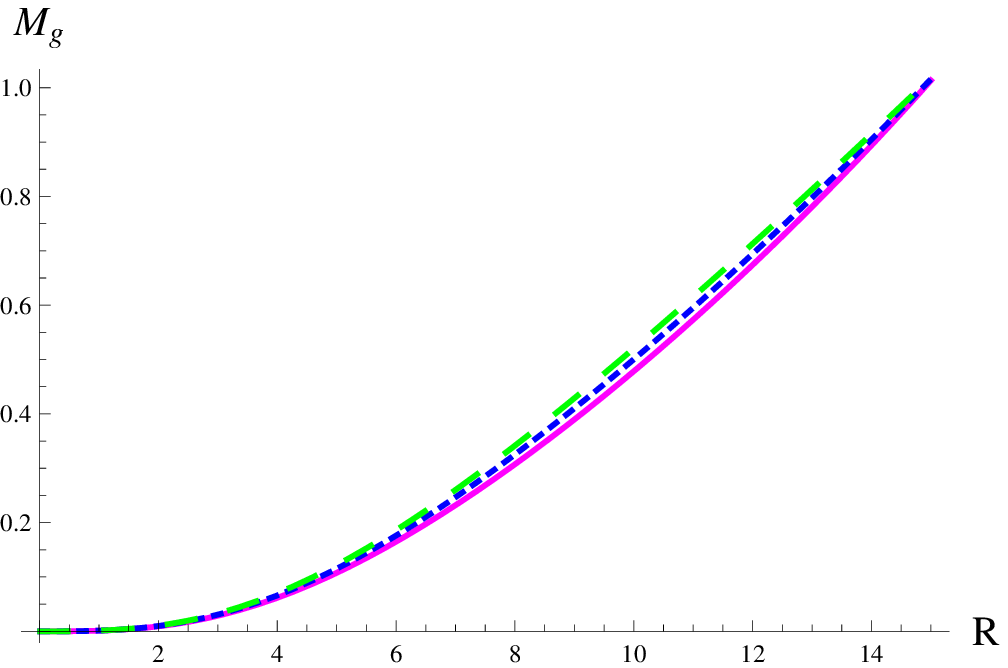,width=0.35\linewidth}\epsfig{file=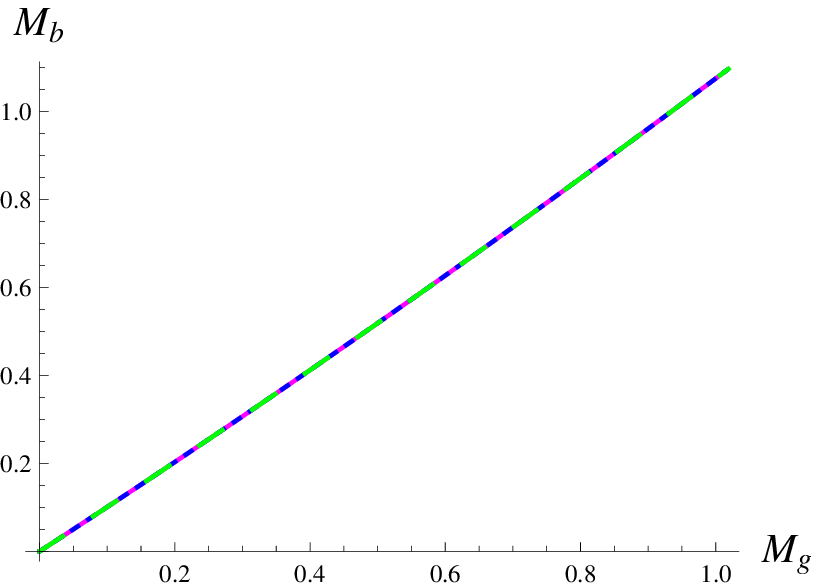,width=0.35\linewidth}
\caption{Plots of gravitational mass versus radius (left) and
baryonic mass versus gravitational mass (right) corresponding to
anisotropic Tolman IV with case I.}
\end{figure}
\begin{figure}\center
\epsfig{file=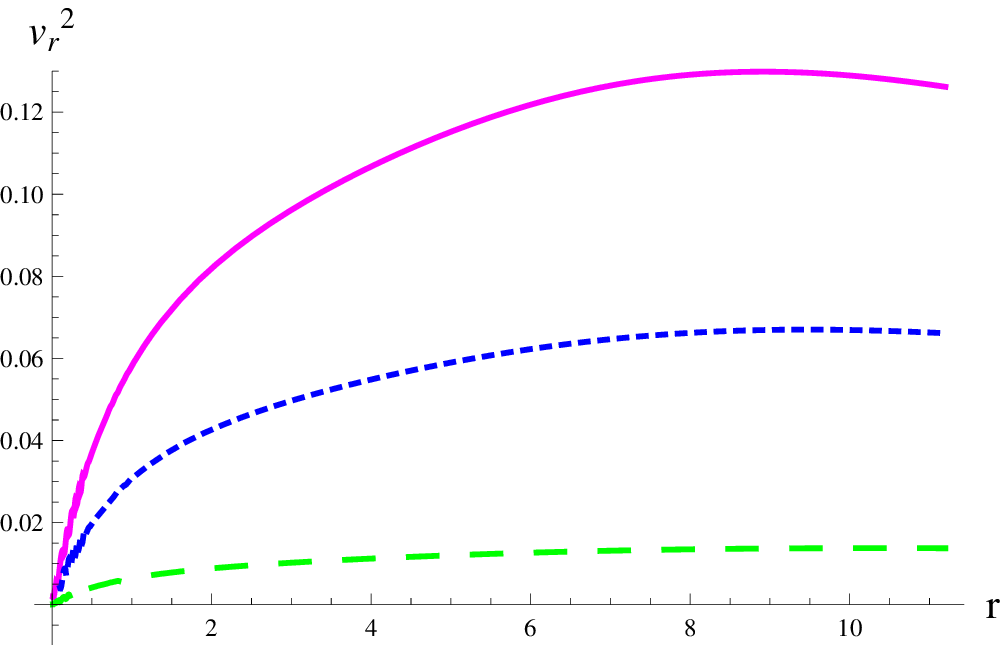,width=0.35\linewidth}\epsfig{file=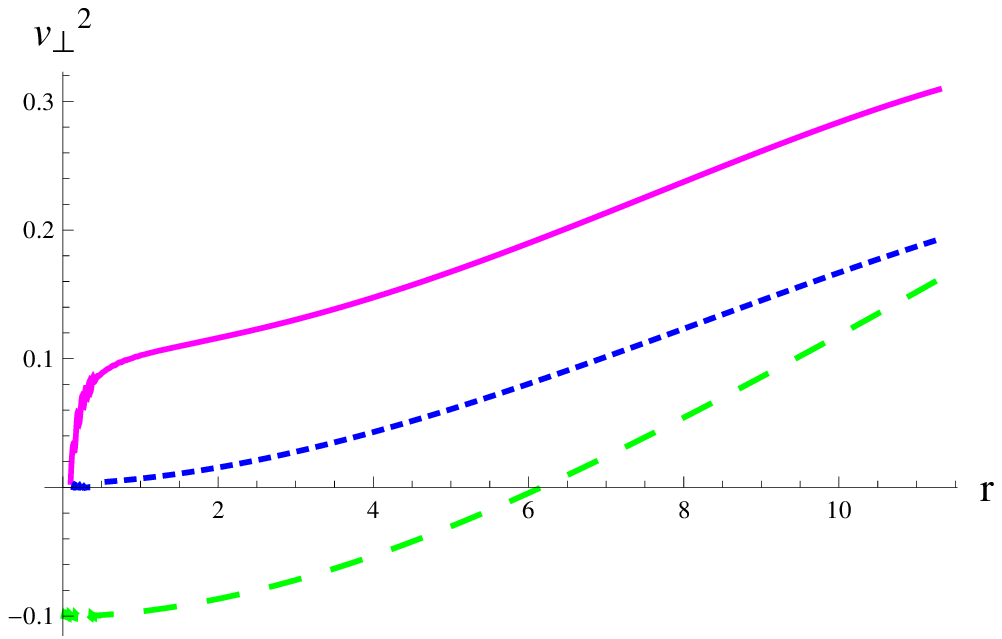,width=0.35\linewidth}\epsfig{file=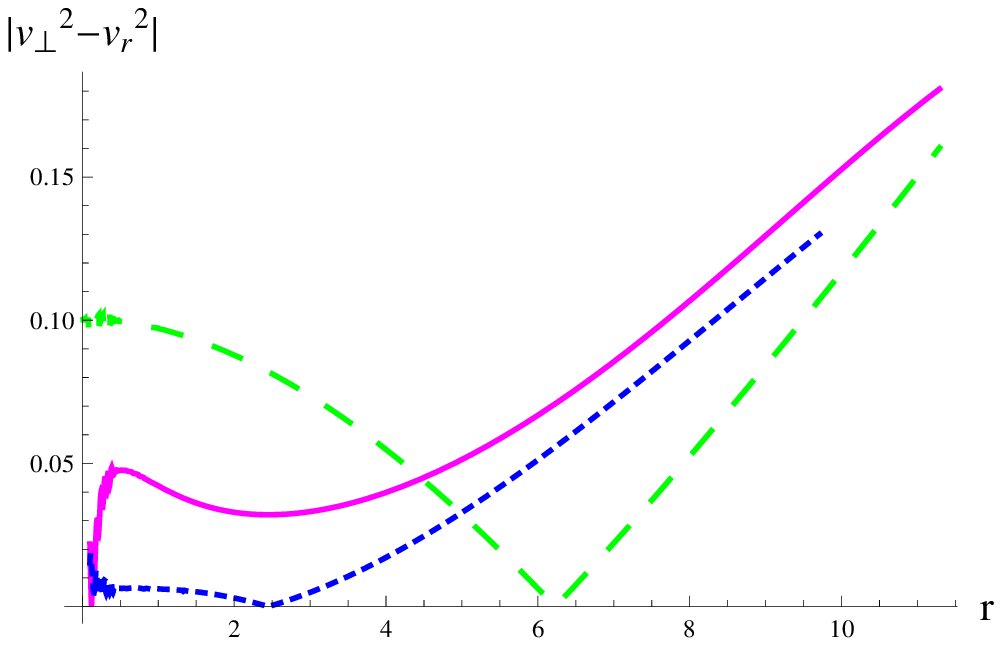,width=0.35\linewidth}
\caption{Plots of radial/tangential velocities and
$|v_\perp^2-v_r^2|$ corresponding to anisotropic Tolman IV for case
I.}
\end{figure}

The internal structure of compact objects is determined by the
gravitational ($M_g$) as well as baryonic $(M_b)$ mass. The
gravitational mass of a spherical gravitationally bound system is
measured using Kepler's law (when a satellite orbits the star) and
is defined as
\begin{equation}\label{50}
M_g=\frac{1}{2}\int_0^R\rho r^2dr.
\end{equation}
The gravitational mass associated with the anisotropic star is
obtained by numerically solving the above equation along with the
Eqs.(\ref{2*}) and (\ref{30''}) under the condition $M_g(0)=0$. The
mass is presented in Figure \textbf{4} as a function of radius for
chosen values of $\alpha$. It is noted that the gravitational mass
of the spherical system increases with an increase in the decoupling
parameter. On the other hand, baryonic mass is directly related to
the massive iron core at the center of the stellar remnant and is
defined as the volume integral of baryon number density times mass
of a baryon. Burrows and Lattimer \cite{51} provided the relation
between gravitational and baryonic mass as
\begin{equation}\label{51}
M_b=M_g+\varpi M_g^2,
\end{equation}
where $\varpi=0.075$ for a large number of nuclear EoS \cite{52}.
The relation between gravitational and baryonic masses, presented in
Figure \textbf{4}, shows that maximum baryonic mass is attained for
$\alpha=0.9$.

The stability of the constructed model is investigated through
causality condition which states that the speed of a propagating
wave is always less than the speed of light \cite{53}. Thus,
according to this criterion the radial $(v_r^2=\frac{dp_r}{d\rho})$
and tangential $(v_\perp^2=\frac{dp_\perp}{d\rho})$ components of
sound speed must lie within the interval $(0,1)$. The plots in
Figure \textbf{5} clearly show that the anisotropic model is stable
for $\alpha=0.2,~0.55$ whereas tangential velocity becomes positive
after a certain distance corresponding to $\alpha=0.9$. Herrera's
cracking approach \cite{54} is another method for determining the
stability of the stellar model. A system is stable with respect to
this concept if the inward directed radial forces of a perturbed
system maintain the same direction throughout the setup, i.e., a
region is stable if radial/tangential components of velocity satisfy
the relation $0<|v_\perp^2-v_r^2|<1$. The extended Tolman IV
solution complies with this condition for $\alpha=0.2,~0.55$ as
shown in Figure \textbf{5}.
\begin{figure}\center
\epsfig{file=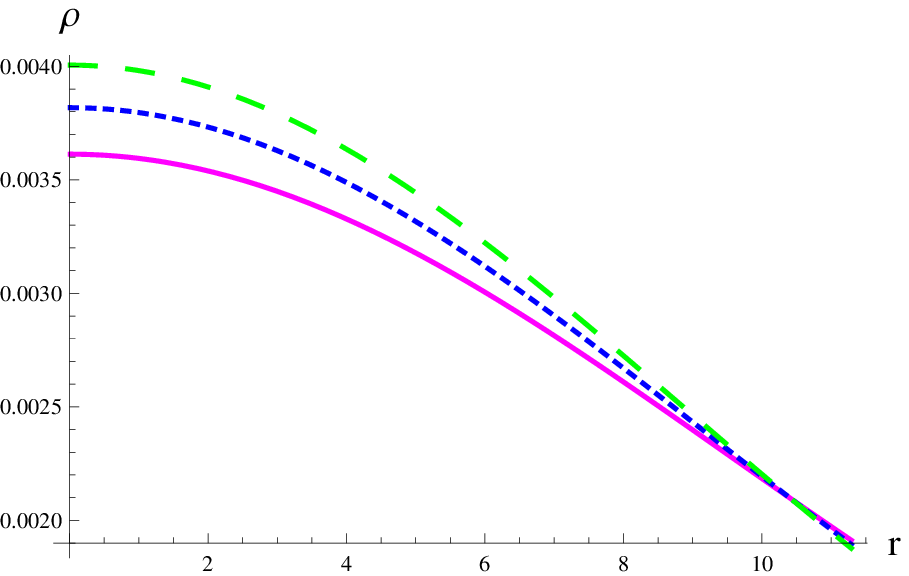,width=0.4\linewidth}\epsfig{file=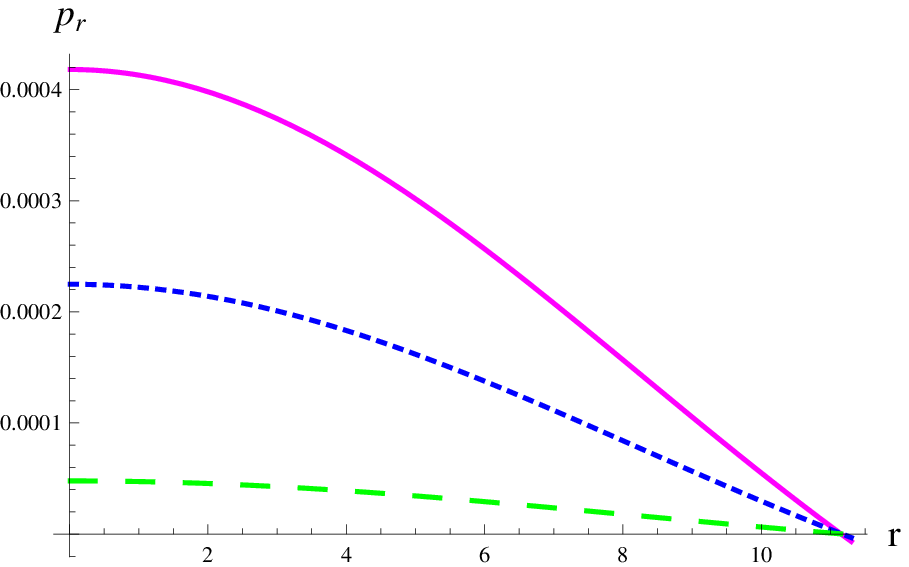,width=0.4\linewidth}
\epsfig{file=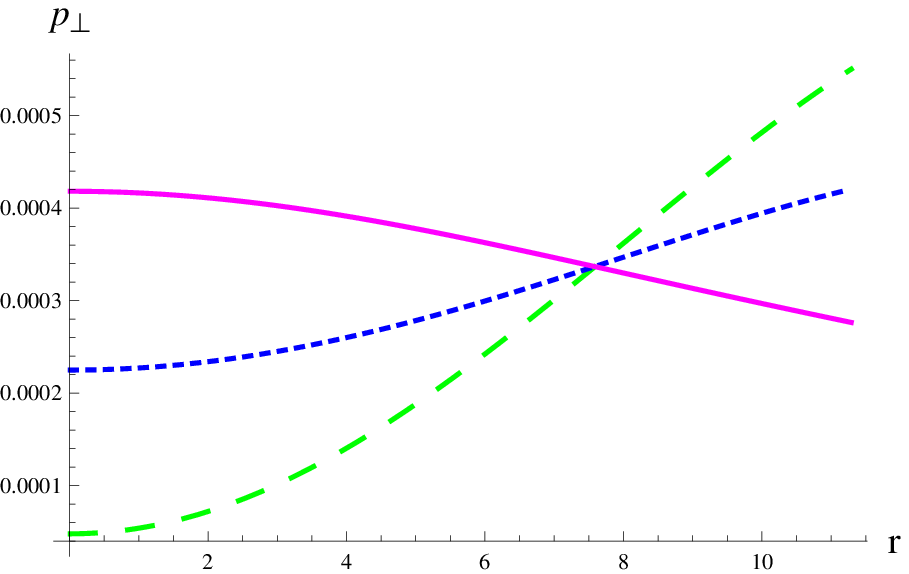,width=0.4\linewidth}\epsfig{file=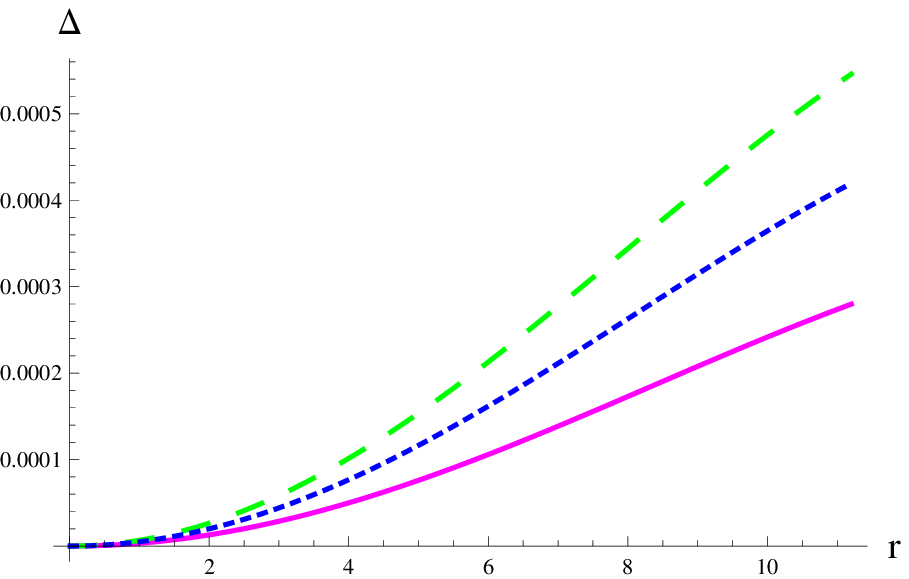,width=0.4\linewidth}
\caption{Plots of matter variables and anisotropy of extended KB
solution for case I.}
\end{figure}

The anisotropic version of the KB solution is formulated in SBD
gravity through Eqs.(\ref{18'}), (\ref{28})-(\ref{30}), (\ref{30''})
and (\ref{30'}). Plots of state variables are presented in Figure
\textbf{6} for $\omega_{BD}=9.87$. The profiles of energy density
and pressure components attain maximum value at the center and
decrease towards the surface for $\alpha=0.2$. However, for higher
values of the decoupling parameter (0.55, 0.9), tangential pressure
exhibits monotonically increasing behavior. Furthermore,  the
anisotropy vanishes at the center as required. This anisotropic
solution is consistent with all energy bounds for chosen values of
$\alpha$ (Figure \textbf{7}) leading to a viable configuration. The
compactness factor and surface redshift obey the desired restraints
as shown in Figure \textbf{8}. Figure \textbf{9} shows an increment
in the gravitational mass as $\alpha$ increases from 0.2 to 0.55.
However, a drop in the mass is observed for a higher value of
$\alpha$. Moreover, the baryonic mass is maximum for $\alpha=0.55$.
The anisotropic model violates the causality condition as tangential
velocity is negative throughout the system for selected values of
$\alpha$ (refer to Figure \textbf{10}). However, the compact object
is stable with respect to Herrera's cracking approach.
\begin{figure}\center
\epsfig{file=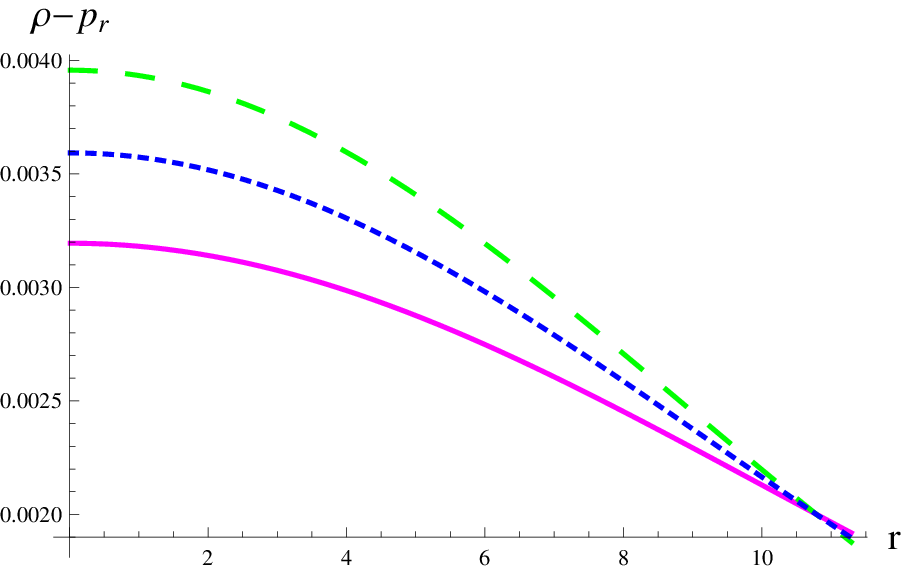,width=0.35\linewidth}\epsfig{file=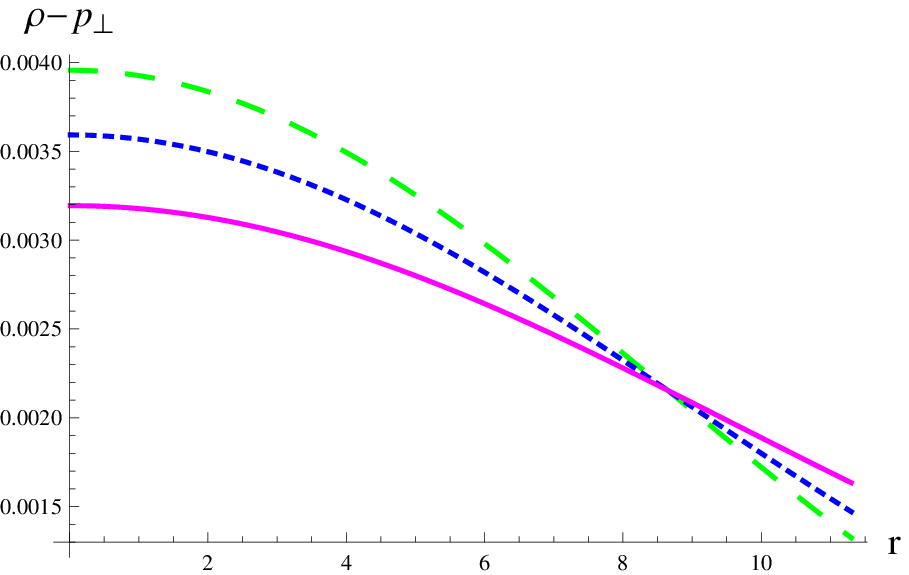,width=0.35\linewidth}
\caption{DEC for extended KB solution with case I.}
\end{figure}
\begin{figure}\center
\epsfig{file=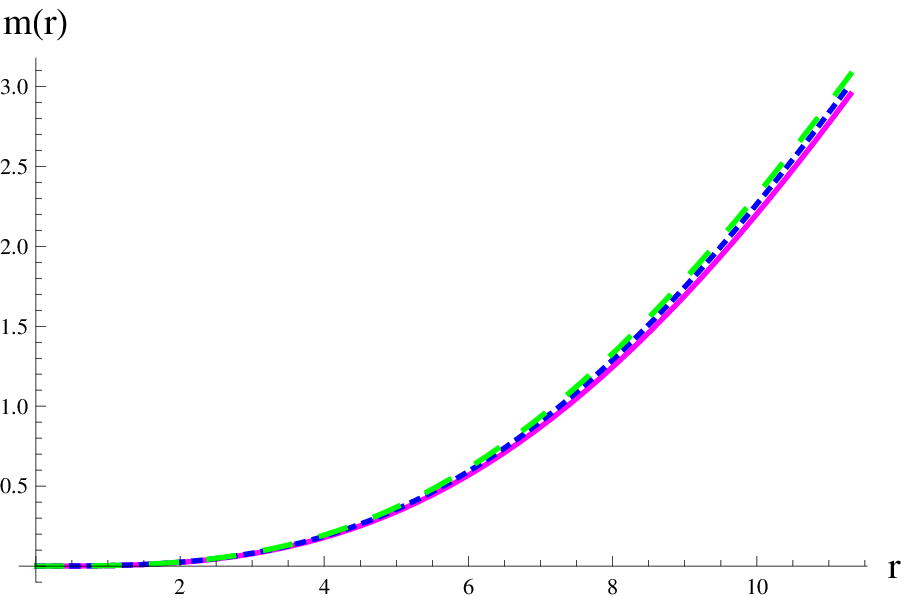,width=0.35\linewidth}\epsfig{file=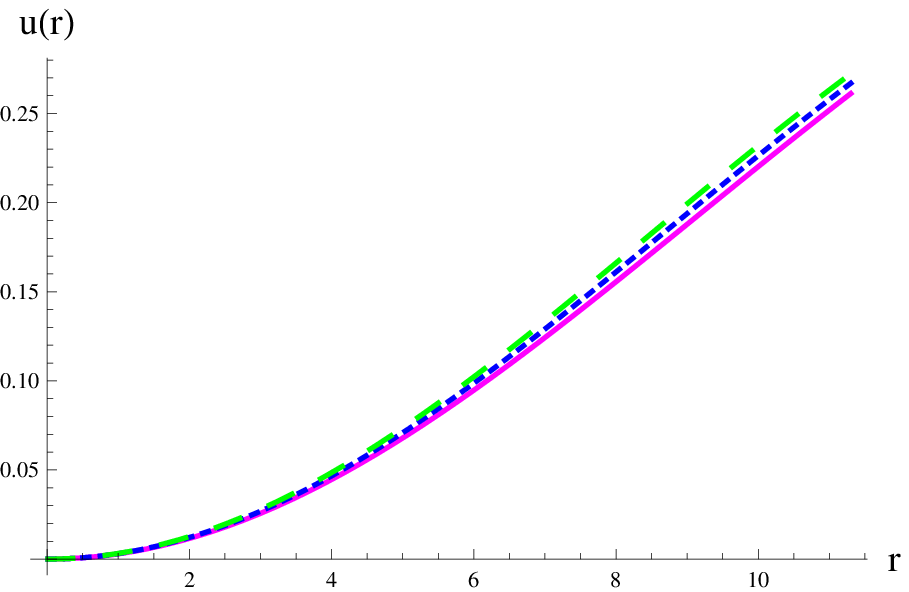,width=0.35\linewidth}\epsfig{file=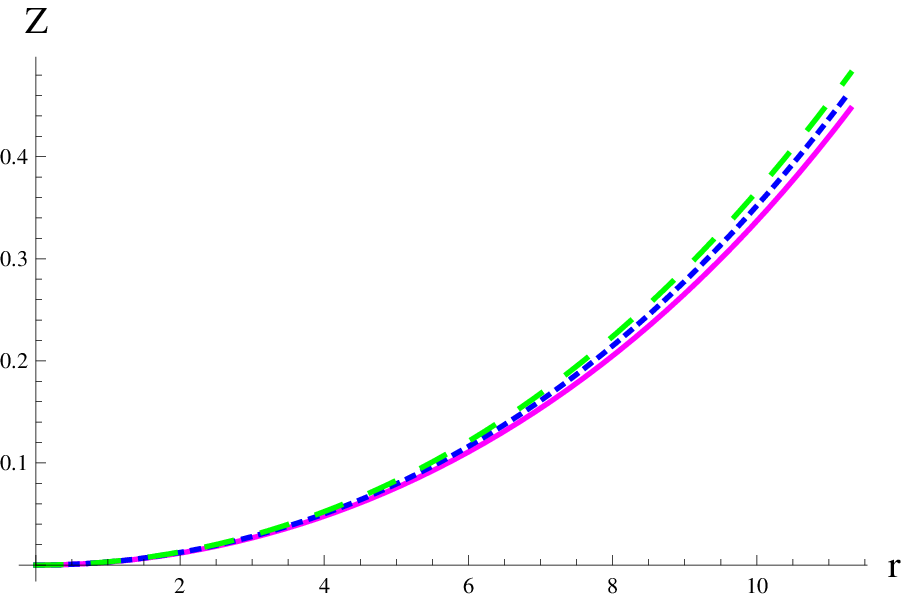,width=0.35\linewidth}
\caption{Plots of mass, compactness and redshift parameters
corresponding to extended KB solution for case I.}
\end{figure}
\begin{figure}\center
\epsfig{file=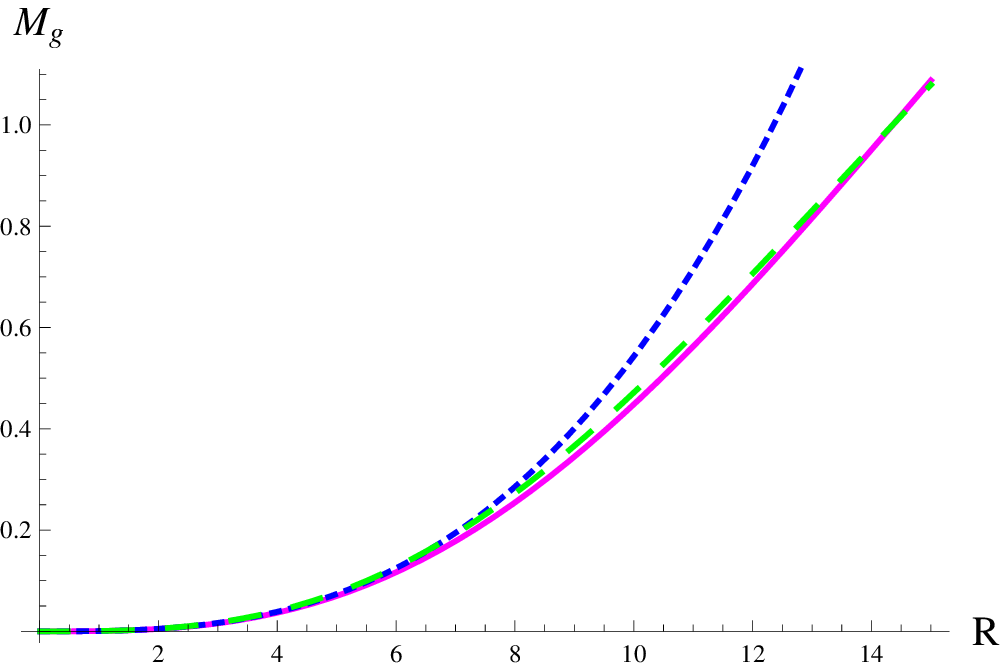,width=0.35\linewidth}\epsfig{file=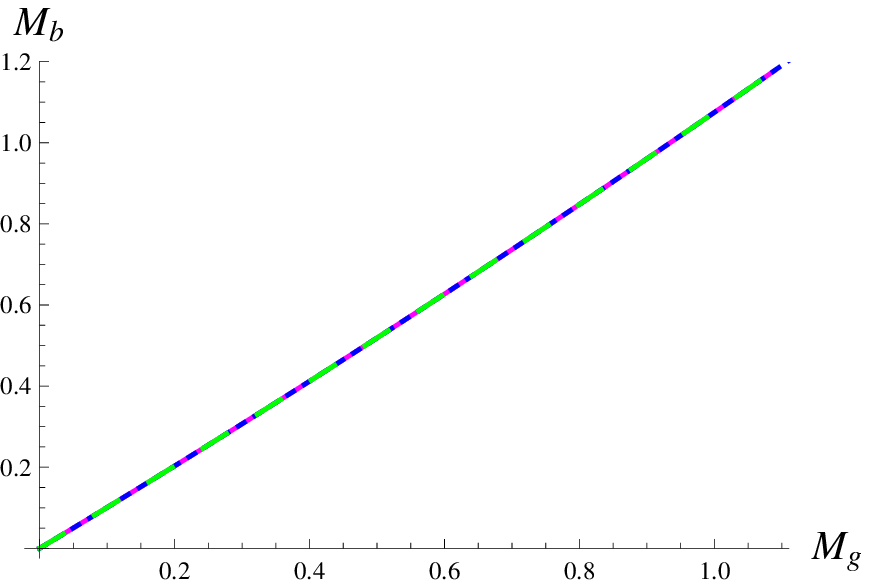,width=0.35\linewidth}
\caption{Plots of gravitational mass versus radius (left) and
baryonic mass versus gravitational mass (right) for extended KB
solution with case I.}
\end{figure}
\begin{figure}\center
\epsfig{file=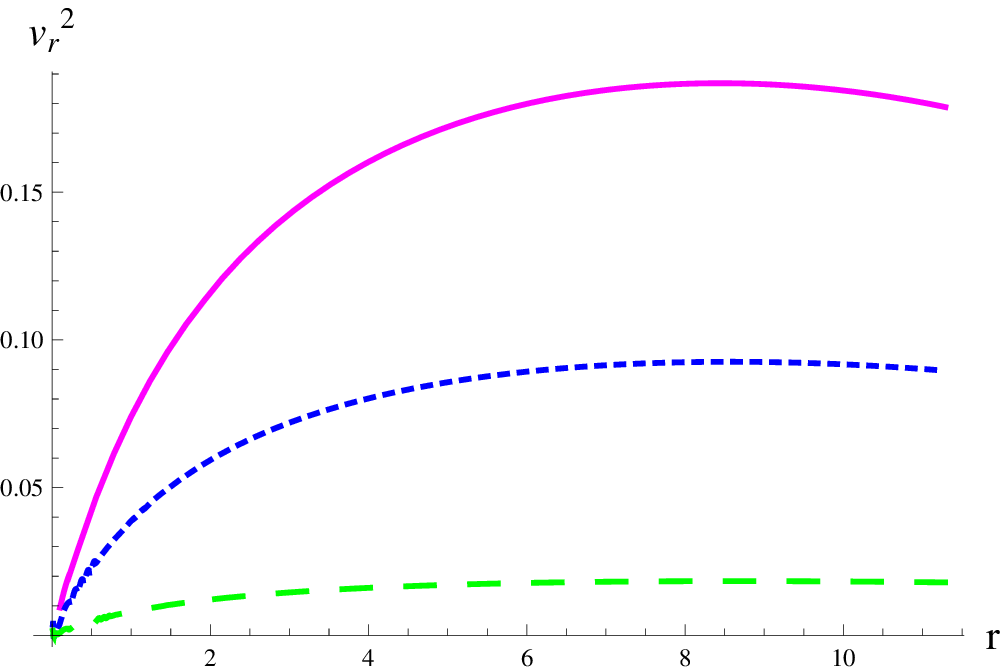,width=0.35\linewidth}\epsfig{file=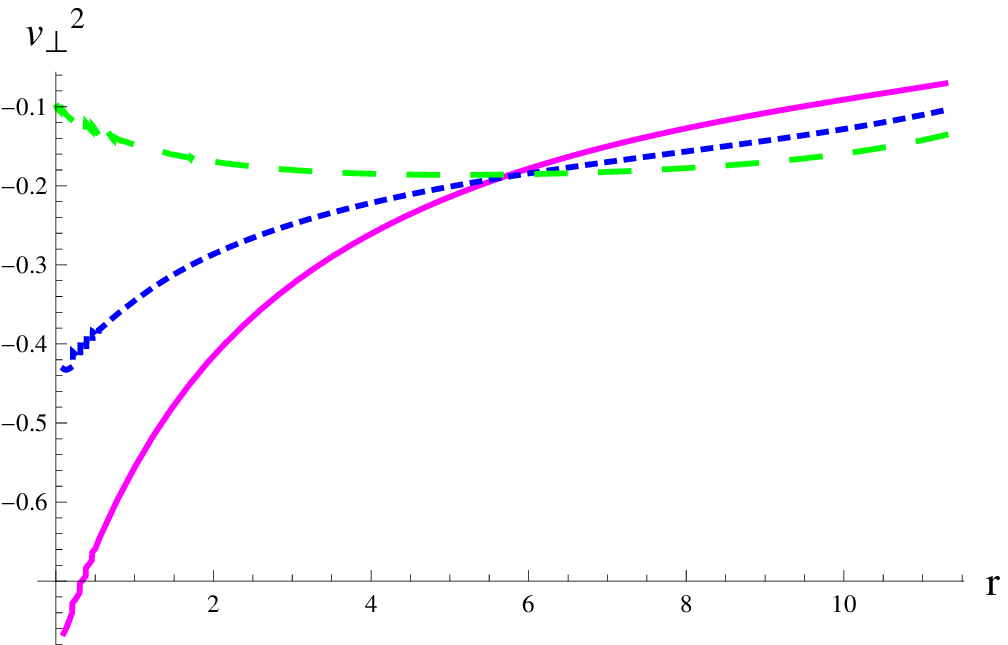,width=0.35\linewidth}\epsfig{file=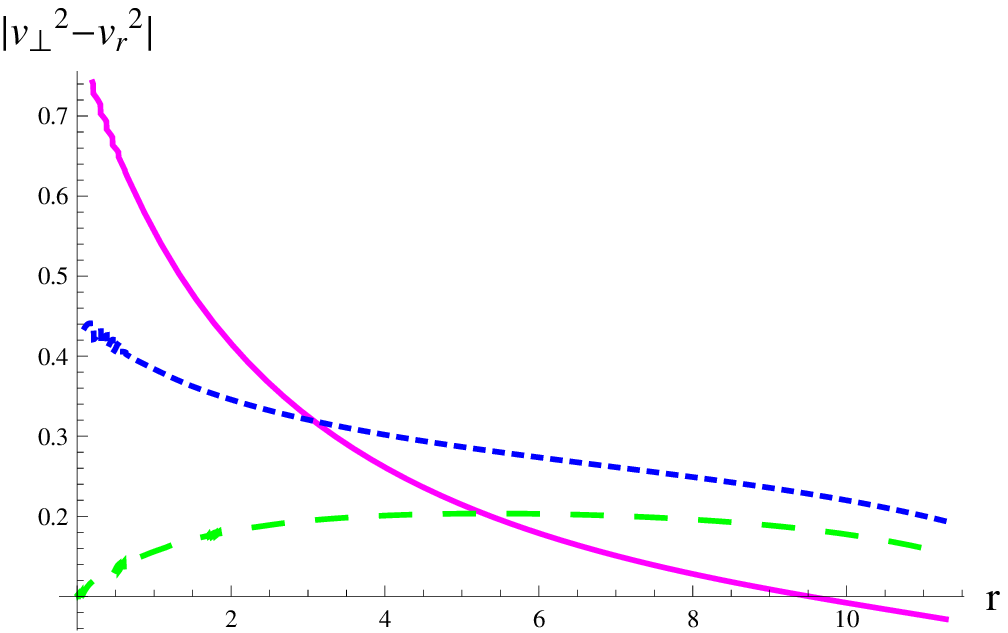,width=0.35\linewidth}
\caption{Plots of radial/tangential velocities and
$|v_\perp^2-v_r^2|$ for extended KB solution with case I.}
\end{figure}

\subsection{Case II: Regularity Condition on Anisotropy}

Bowers and Liang \cite{35} proposed that singularities in the
Tolman-Oppenheimer-Volkoff equation can be avoided if the following
condition is imposed on the anisotropy
\begin{equation*}
p_\perp-p_r=Ch(p_r,r)(\rho+p_r)r^n,
\end{equation*}
where the parameter $C$ measures the strength of the anisotropy and
$n>1$. For the present work, we have taken $C=-0.5$. Moreover, $h$
is an arbitrary function of radial pressure and contains information
about the anisotropy of the system. In 1981, Cosenza et al.
\cite{36} evaluated anisotropic solutions from known isotropic
solutions by assuming the energy density of a perfect fluid and
taking $h(p_r,r)=\frac{\upsilon'(r)}{2}r^{1-n}$. These conditions
have already been employed in MGD approach to obtain new anisotropic
solutions \cite{37, 50}. In this section, we obtain anisotropic
analogues of seed solutions by imposing Bowers-Liang constraint on
$\Theta$-sector as
\begin{equation}\label{31'}
\Theta^2_2-\Theta^1_1=Ch(\Theta^1_1,r)(-\Theta^0_0+\Theta^1_1)r^n,
\end{equation}
with $h(\Theta^1_1,r)=\frac{\upsilon'(r)}{2}r^{1-n}$. Substituting
Eqs.(\ref{15})-(\ref{17}) in the above equation leads to
\begin{eqnarray}\nonumber
&&\frac{e^{-\eta(r)}}{r\Psi(r)}\left(f(r)e^{\eta(r)}\left(r\Psi(r)\left(\Psi'(r)\left(C\alpha
r^2g'(r)\upsilon'(r)+Cr^2\mu'(r)\upsilon'(r)+4\right)\right.\right.\right.\\\nonumber
&&-\left.\left.\left.2r\Psi''(r)\left(Cr\upsilon'(r)+2\right)\right)+
\Psi^2(r)\left(-\left(2\left(\alpha
r^2g''(r)+r^2\mu''(r)-2\right)\right.\right.\right.\right.\\\nonumber
&&+\left.\left.\left.\left.\alpha^2r^2g'^2(r)+2\alpha
rg'(r)\left(r\mu'(r)-Cr\upsilon'(r)-1\right)+r^2\mu'^2(r)-2r\mu'(r)\right.\right.\right.\right.\\\nonumber
&&\times\left.\left.\left.\left.\left(Cr\upsilon'(r)+1\right)\right)\right)-2r^2\omega_{BD}\Psi'^2(r)
\left(Cr\upsilon'(r)+2\right)\right)-r\Psi(r)\left(r\Psi'(r)\right.\right.\\\nonumber
&&\times\left.\left.\left(e^{\eta(r)}
f'(r)\left(Cr\upsilon'(r)+2\right)-Crg'(r)\upsilon'(r)\right)+\Psi(r)\left(e^{\eta
(r)}f'(r)\left(\alpha rg'(r)\right.\right.\right.\right.\\\nonumber
&&+\left.\left.\left.\left.r\mu'(r)+2Cr\upsilon'(r)+2\right)+2r
g''(r)+\alpha
rg'^2(r)+g'(r)\left(2r\mu'(r)-r\eta'(r)\right.\right.\right.\right.\\\label{33}
&&-\left.\left.\left.\left.2Cr\upsilon
'(r)-2\right)\right)\right)\right)=0.
\end{eqnarray}

We obtain the deformation function $g(r)$ by simultaneously solving
Eqs.(\ref{2}) and (\ref{33}) numerically with the initial conditions
$\Psi(0)=0.1,~\Psi'(0)=0,~g(0)=0$ and $g'(0)=0.5$. The function
$f(r)$ is evaluated from the constraint (\ref{27}) as
\begin{eqnarray}\nonumber
f(r)&=&e^{-\eta(r)}\left(2\Psi^2(r)\left(-rg'(r)-r\mu
'(r)+e^{\eta(r)}-1\right)-r\Psi(r)\Psi'(r)\left(rg'(r)\right.\right.\\\nonumber
&+&\left.\left.r\mu'(r)+4\right)+r^2e^{\eta(r)}\Psi(r)V(\Psi)+r^2\omega_{BD}\Psi
'^2(r)\right)(r\Psi(r)\Psi'\left(\alpha
rg'(r)\right.\\\label{32}&+&\left.r\mu'(r)+4\right)+2\Psi^2(r)\left(\alpha
rg'(r)+r\mu '(r)+1\right)-r^2\omega_{BD}\Psi'^2(r))^{-1}.
\end{eqnarray}
Tolman IV solution is extended via constraint (\ref{31'}) by
employing the associated constants in Eqs.(\ref{28})-(\ref{30}),
(\ref{33}) and (\ref{32}). The physical characteristics of this
solution are investigated graphically for $\omega_{BD}=17.95$.
Figure \textbf{11} displays the energy density and pressures as
decreasing functions of $r$ for the considered values of $\alpha$. A
decrease in the physical parameters ($\rho,~p_r,~p_\perp$) is
observed for higher values of the decoupling parameter whereas
anisotropy increases as $\alpha$ increases. Moreover, the anisotropy
within the star increases for some distance and then decreases
indicating the presence of a weaker repulsive force near the stellar
surface. Figure \textbf{12} shows that the system corresponding to
extended Tolman IV solution is viable as it adheres to the
restrictions imposed by energy bounds. Moreover, the compactness and
redshift parameters (Figure \textbf{13}) adhere to the respective
bounds. The gravitational and baryonic masses calculated from
Eqs.(\ref{50}) and (\ref{51}), respectively are plotted in Figure
\textbf{14}. The compact structure becomes more massive as $\alpha$
increases form 0.2 to 0.55 but decreases for $\alpha=0.9$.
Furthermore, the model has maximum baryonic mass for $\alpha=0.55$.
Finally, the extended Tolman IV solution is stable for the
considered values of the decoupling parameter as it complies with
the causality condition and cracking approach as shown in Figure
\textbf{15}.
\begin{figure}\center
\epsfig{file=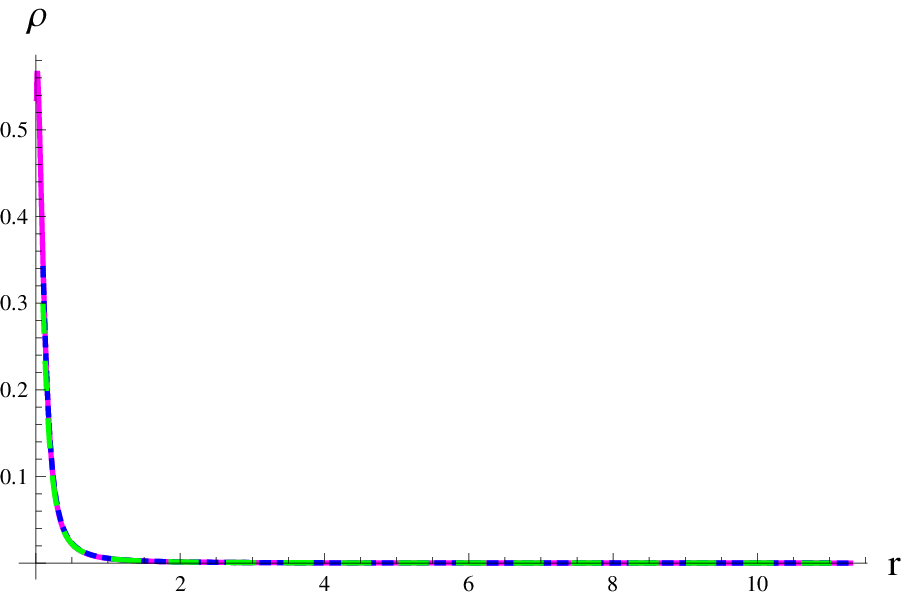,width=0.4\linewidth}\epsfig{file=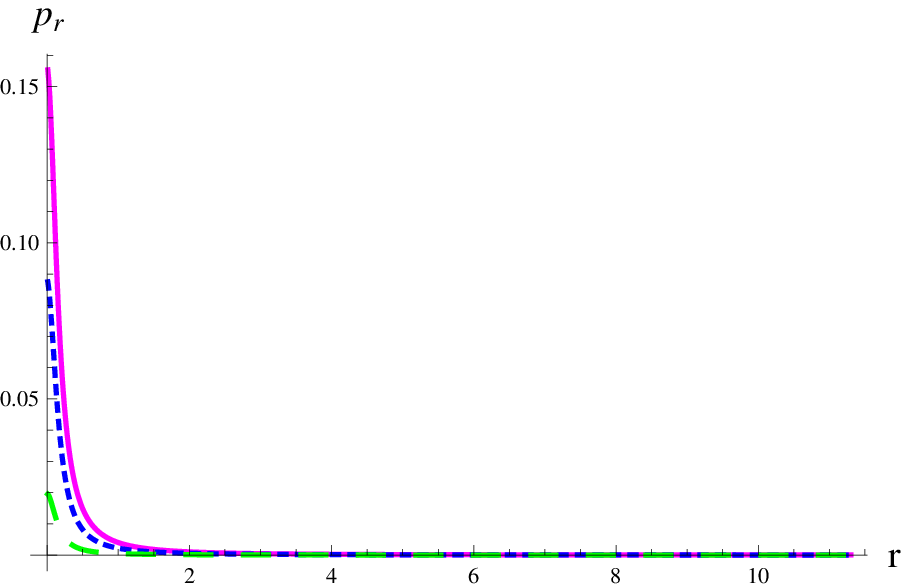,width=0.4\linewidth}
\epsfig{file=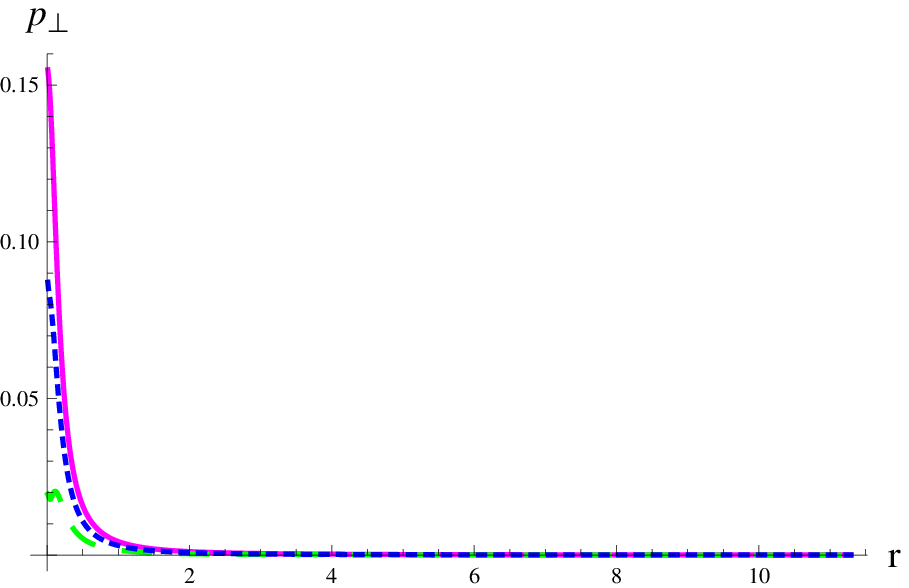,width=0.4\linewidth}\epsfig{file=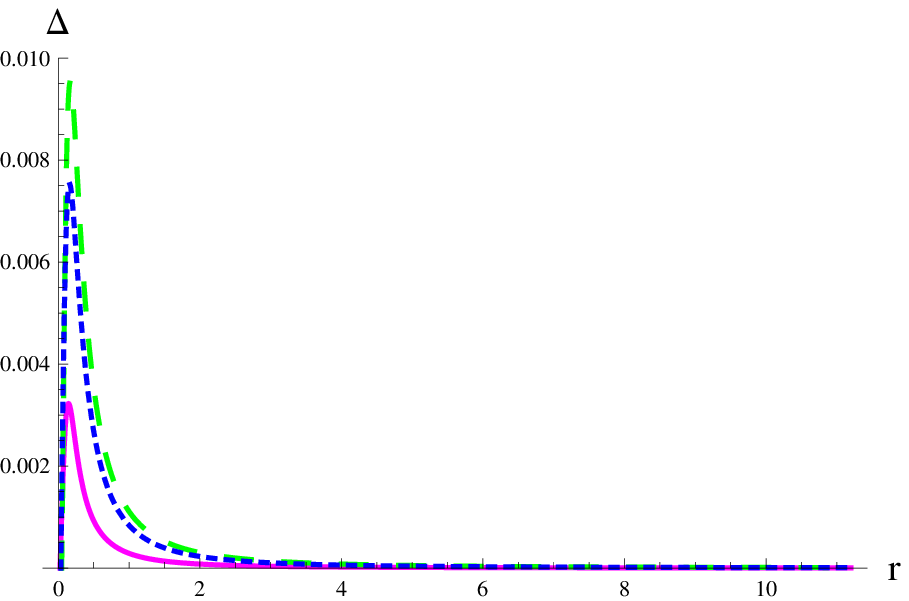,width=0.4\linewidth}
\caption{Plots of matter variables and anisotropy of extended Tolman
IV for case II.}
\end{figure}
\begin{figure}\center
\epsfig{file=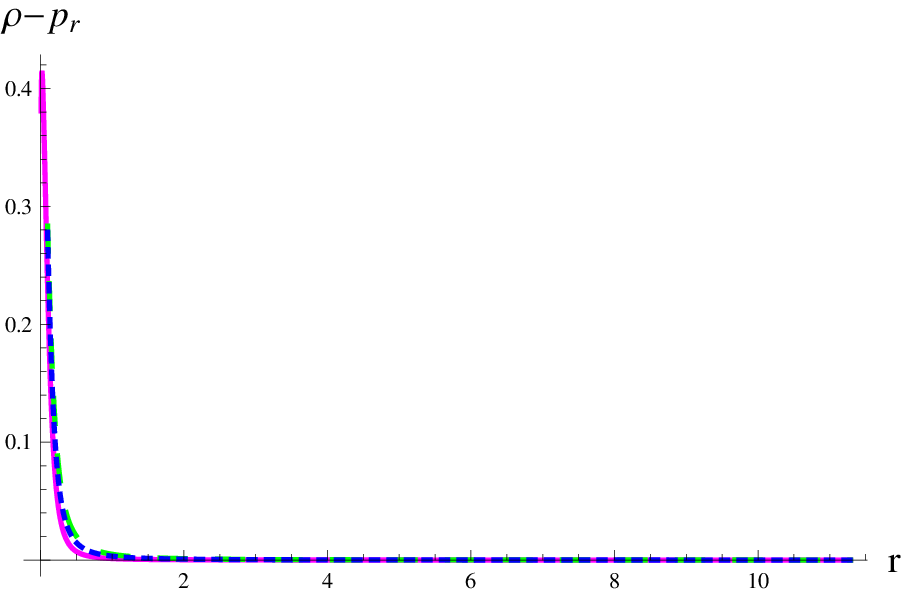,width=0.4\linewidth}\epsfig{file=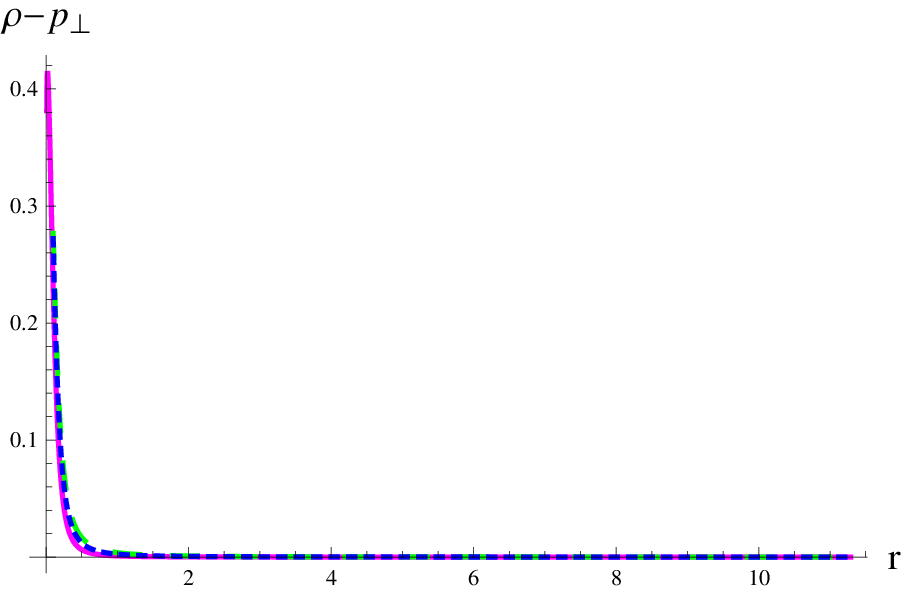,width=0.4\linewidth}
\caption{DEC for anisotropic Tolman IV with case II.}
\end{figure}
\begin{figure}\center
\epsfig{file=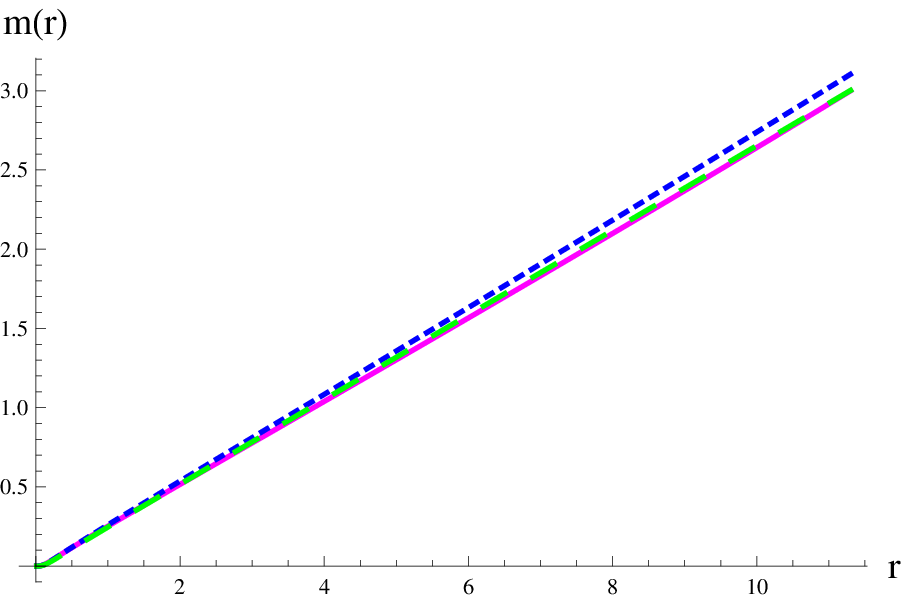,width=0.35\linewidth}\epsfig{file=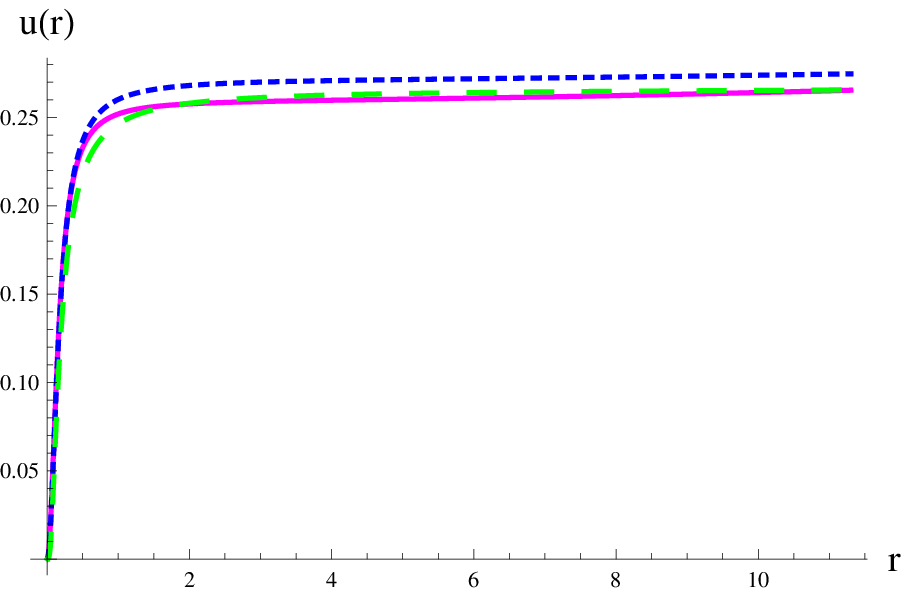,width=0.35\linewidth}\epsfig{file=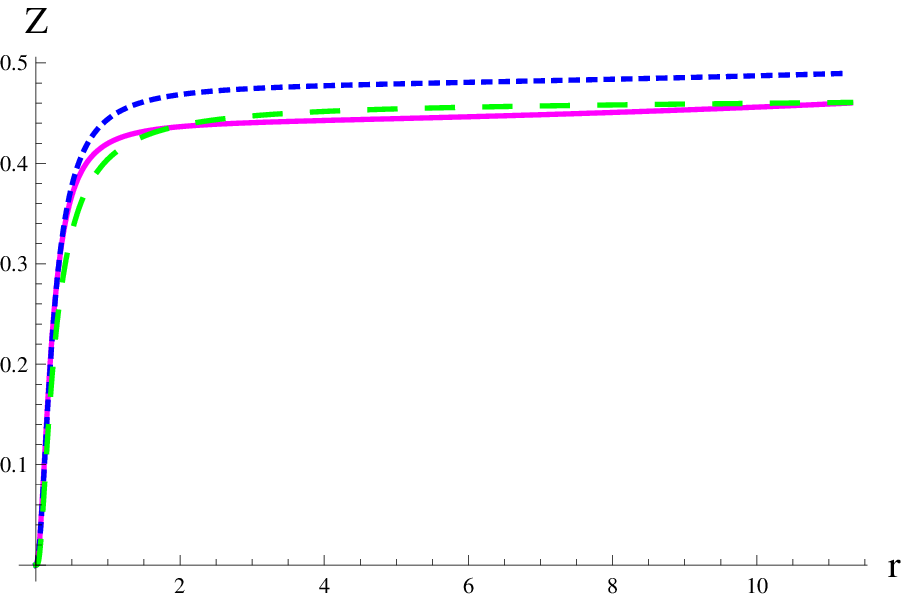,width=0.35\linewidth}
\caption{Plots of mass, compactness and redshift parameters
corresponding to anisotropic Tolman IV for case II.}
\end{figure}
\begin{figure}\center
\epsfig{file=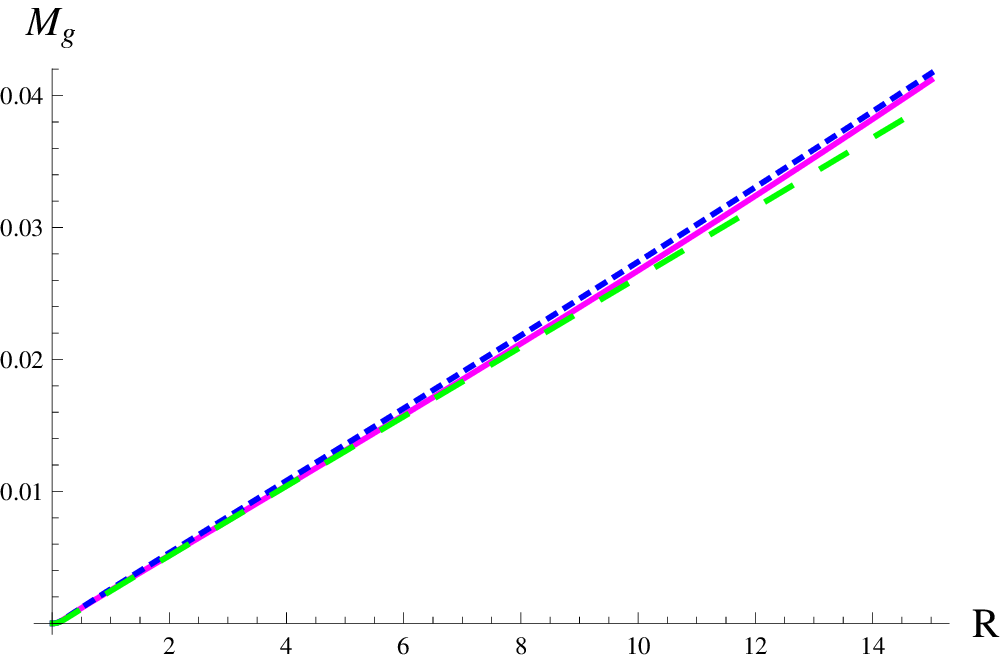,width=0.35\linewidth}\epsfig{file=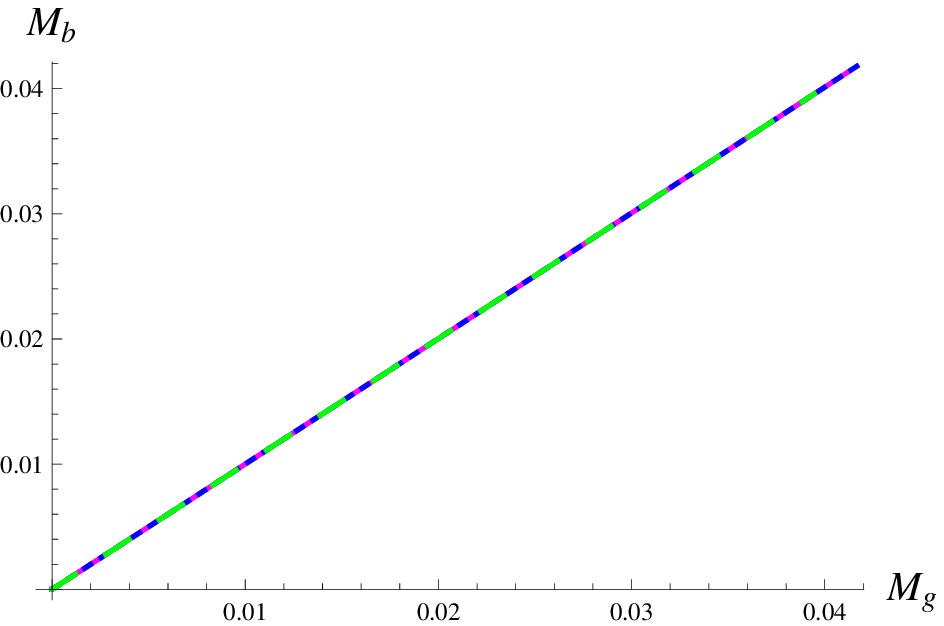,width=0.35\linewidth}
\caption{Plots of gravitational mass versus radius (left) and
baryonic mass versus gravitational mass (right) corresponding to
anisotropic Tolman IV for case II.}
\end{figure}
\begin{figure}\center
\epsfig{file=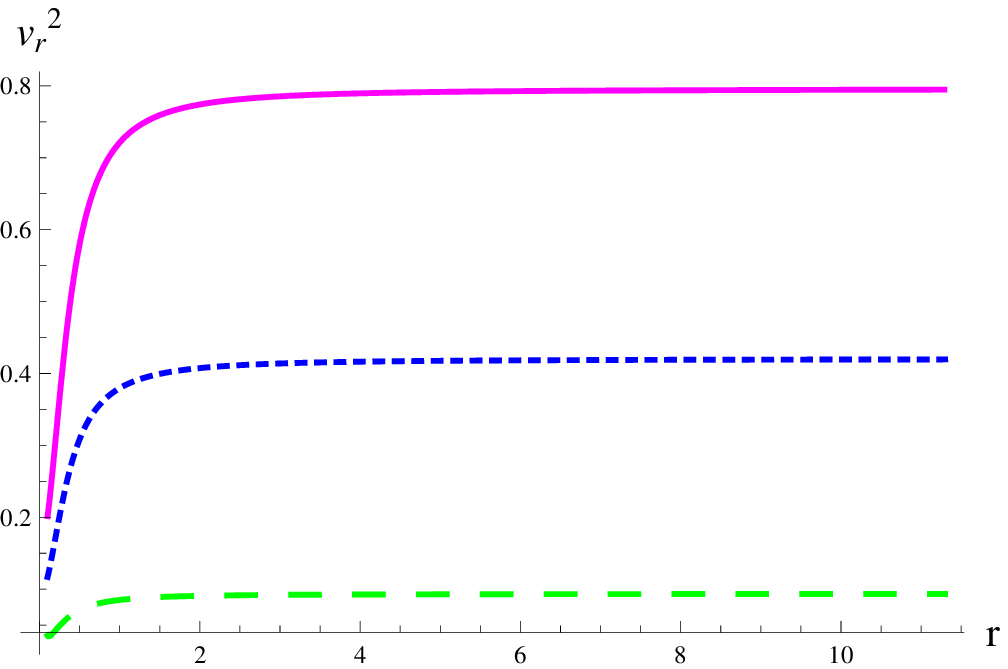,width=0.35\linewidth}\epsfig{file=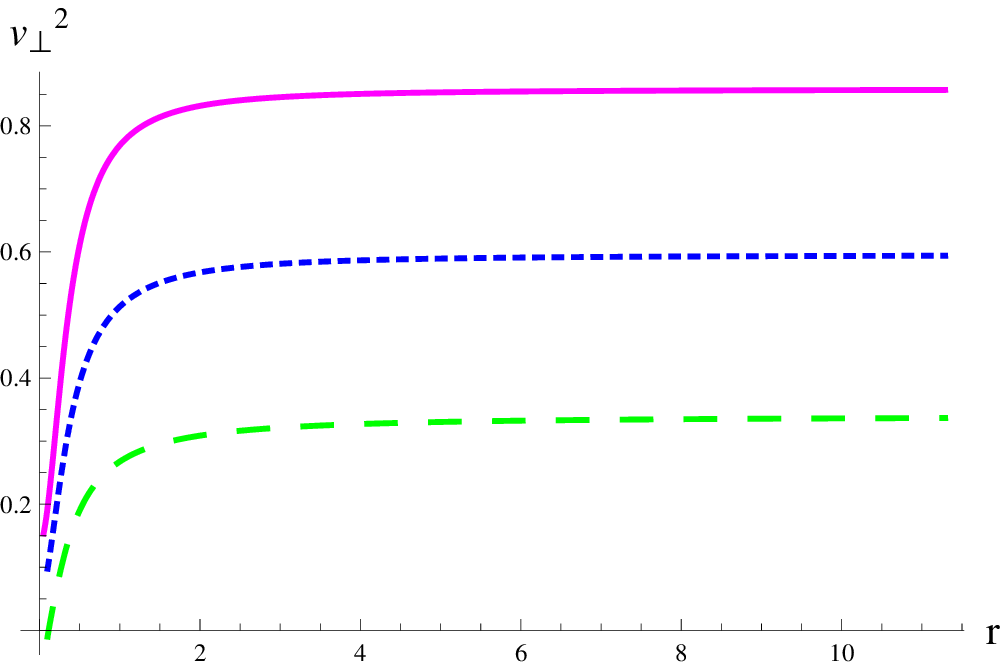,width=0.35\linewidth}\epsfig{file=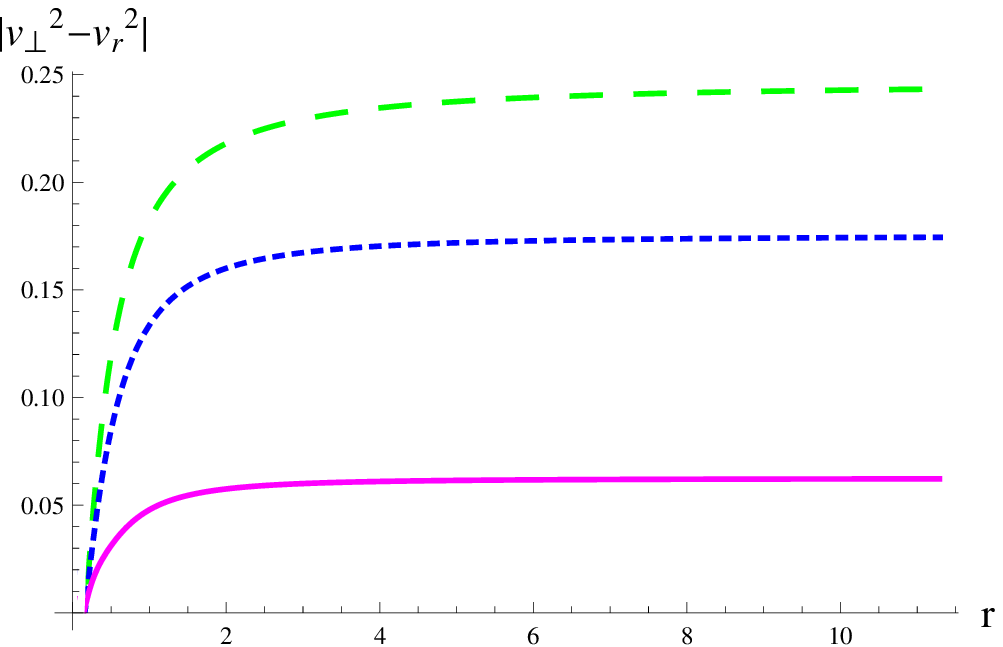,width=0.35\linewidth}
\caption{Plots of radial/tangential velocities and
$|v_\perp^2-v_r^2|$ corresponding to anisotropic Tolman IV for case
II.}
\end{figure}
\begin{figure}\center
\epsfig{file=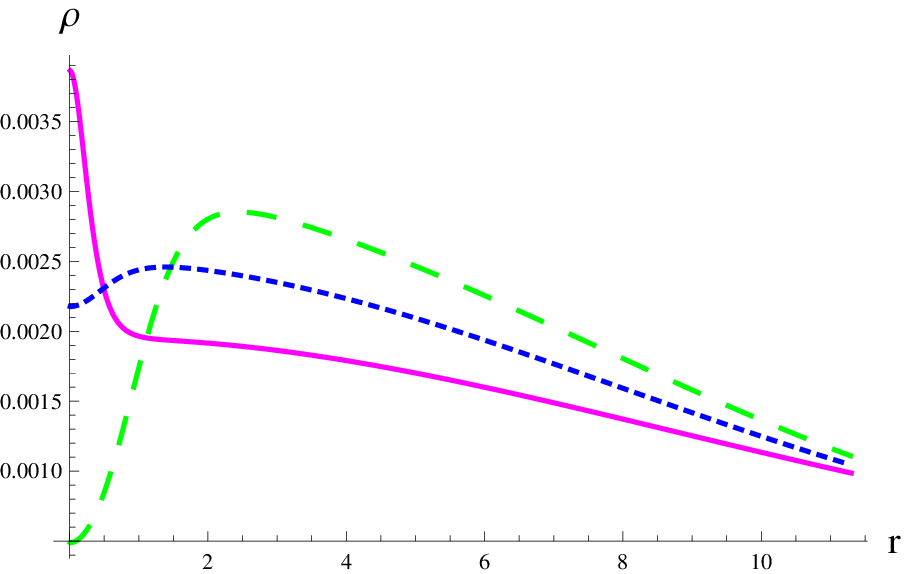,width=0.4\linewidth}\epsfig{file=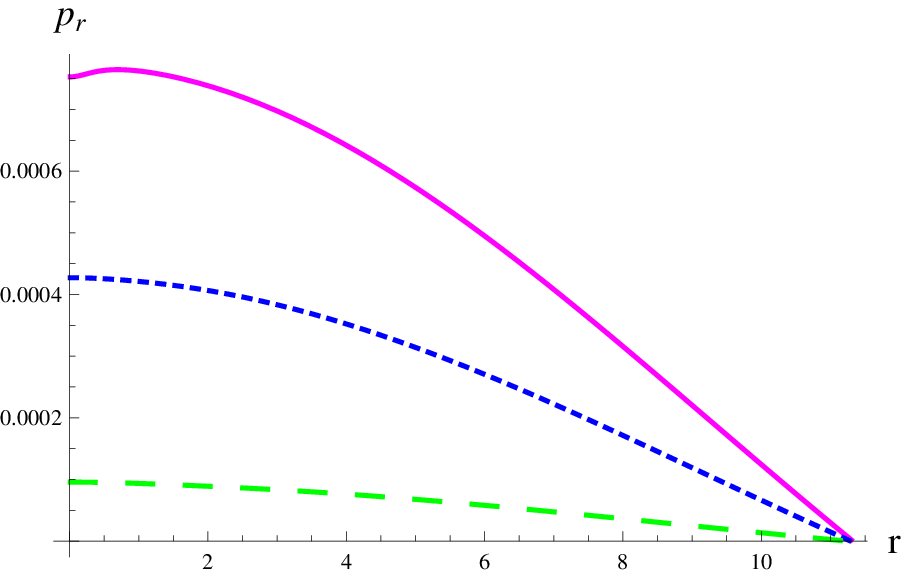,width=0.4\linewidth}
\epsfig{file=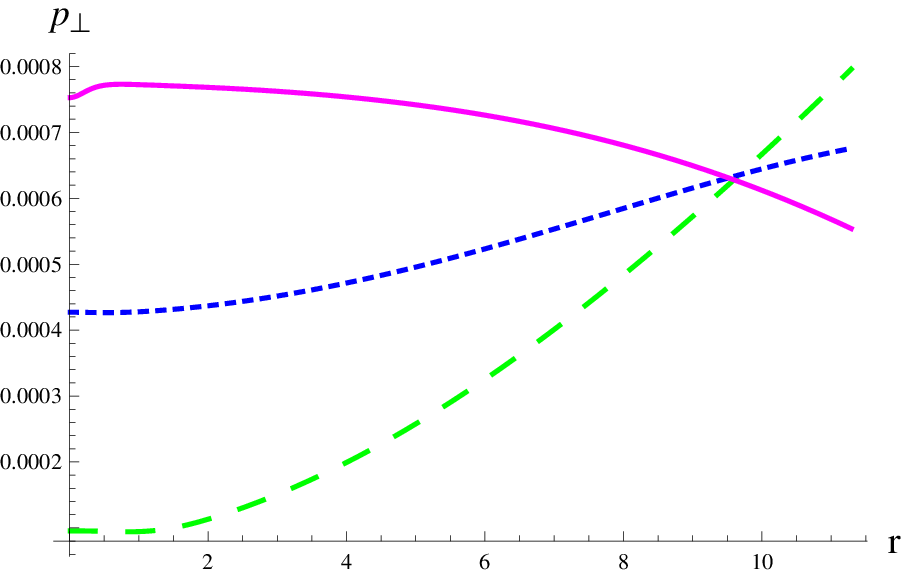,width=0.4\linewidth}\epsfig{file=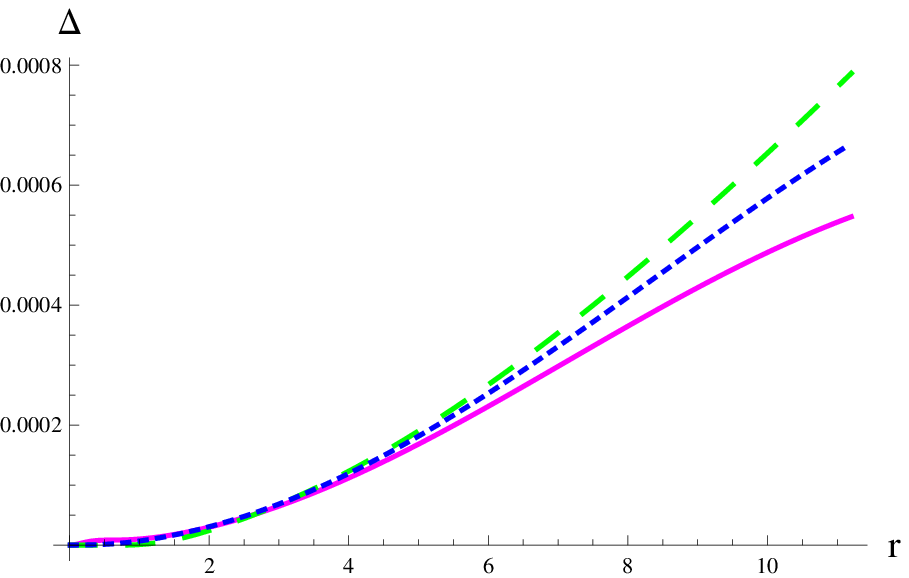,width=0.4\linewidth}
\caption{Plots of matter variables and anisotropy of extended KB
solution for case II.}
\end{figure}
\begin{figure}\center
\epsfig{file=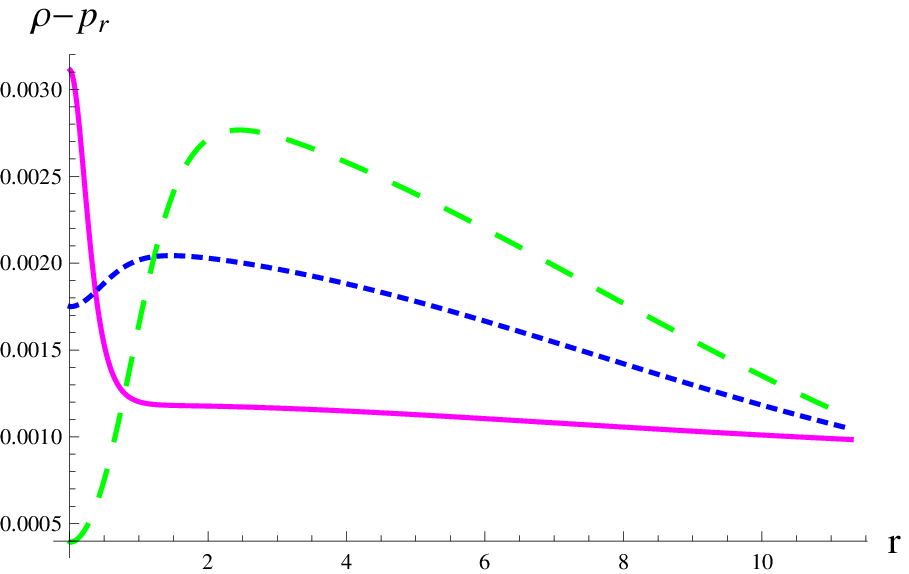,width=0.4\linewidth}\epsfig{file=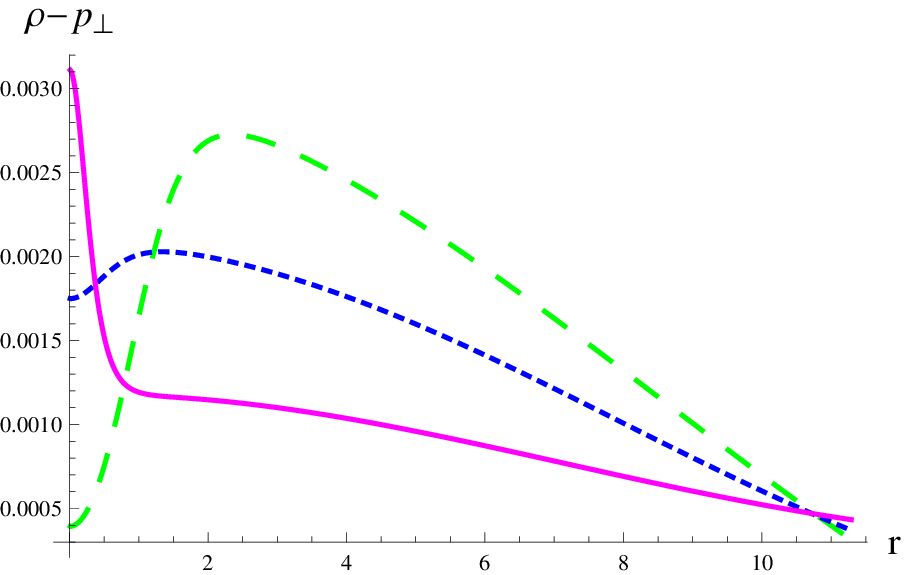,width=0.4\linewidth}
\caption{DEC for extended KB solution with case II.}
\end{figure}

The extended version of KB solution under Bowers-Liang constraint is
constructed by plugging the associated metric potentials and
constants in Eqs.(\ref{28})-(\ref{30}), (\ref{33}) and (\ref{32}).
It is noted from Figure \textbf{16} that energy density and
tangential pressure are positive for the selected values of $\alpha$
but decrease monotonically only for $\alpha=0.2$. However, the
radial pressure has a maximum value at the center and vanishes at
$r=R$ for all values of the decoupling parameter. Moreover, the
state parameters are inversely proportional to $\alpha$ while the
anisotropy is directly proportional to the decoupling parameter. The
plots of DEC in Figure \textbf{17} show that the anisotropic
solution is physically valid for the chosen values of $\alpha$. The
values of compactness parameter and surface redshift also lie in the
desired range as displayed in Figure \textbf{18}. The gravitational
mass increases with increase in the decoupling parameter as shown in
Figure \textbf{19}. Moreover, the baryonic mass is maximum for
$\alpha=0.9$. The constructed model is stable only for $\alpha=0.2$
as it violates causality and cracking conditions for higher values
of the decoupling parameter (Figure \textbf{20}).
\begin{figure}\center
\epsfig{file=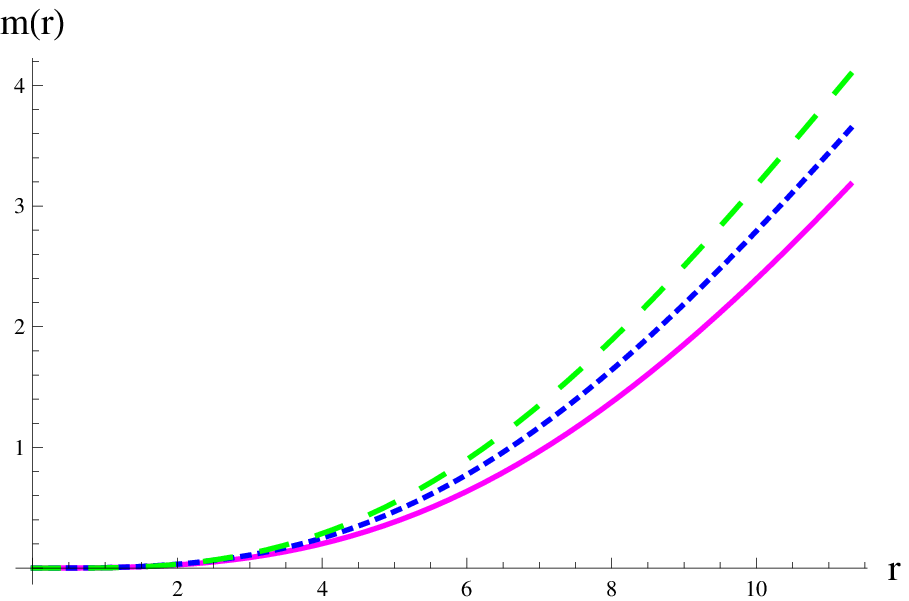,width=0.35\linewidth}\epsfig{file=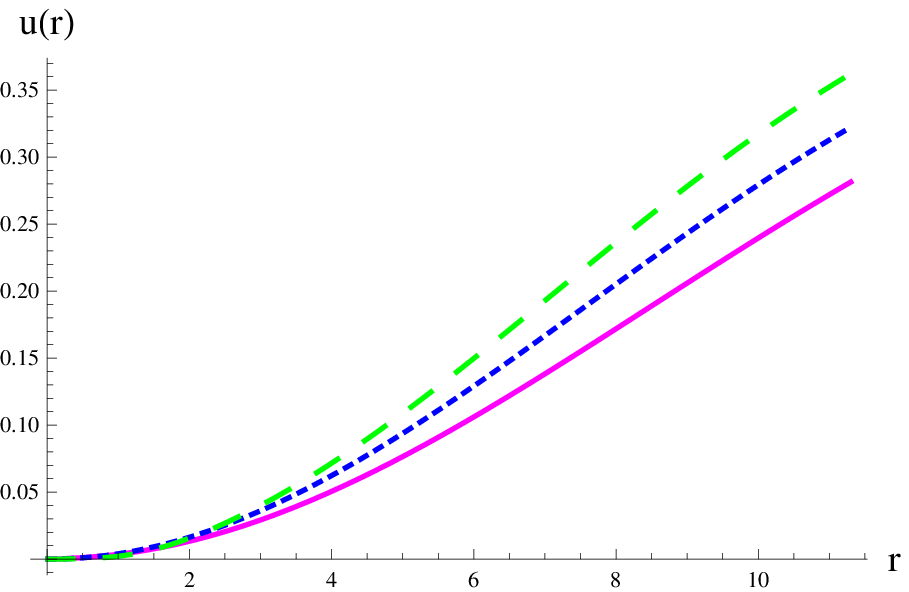,width=0.35\linewidth}\epsfig{file=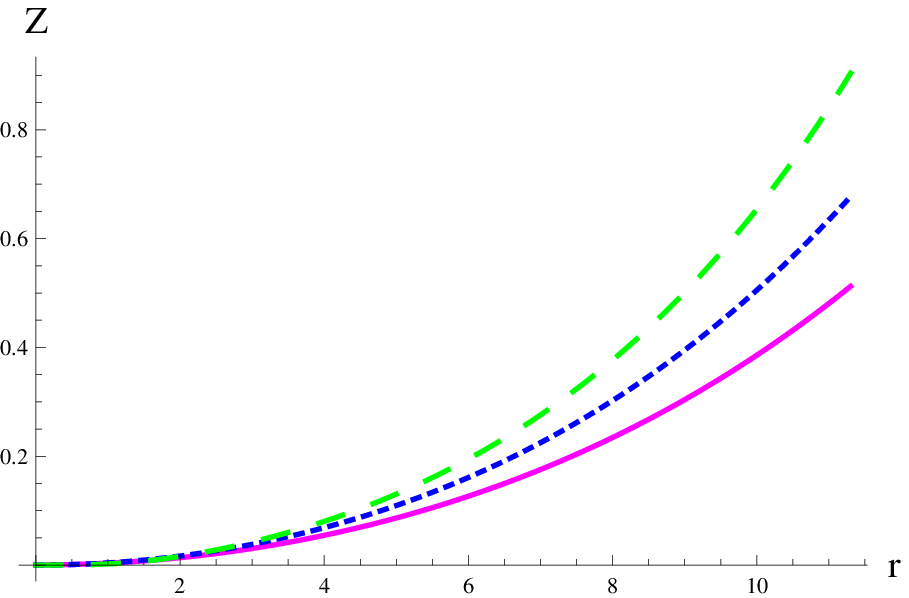,width=0.35\linewidth}
\caption{Plots of mass, compactness and redshift parameters
corresponding to extended KB solution for case II.}
\end{figure}
\begin{figure}\center
\epsfig{file=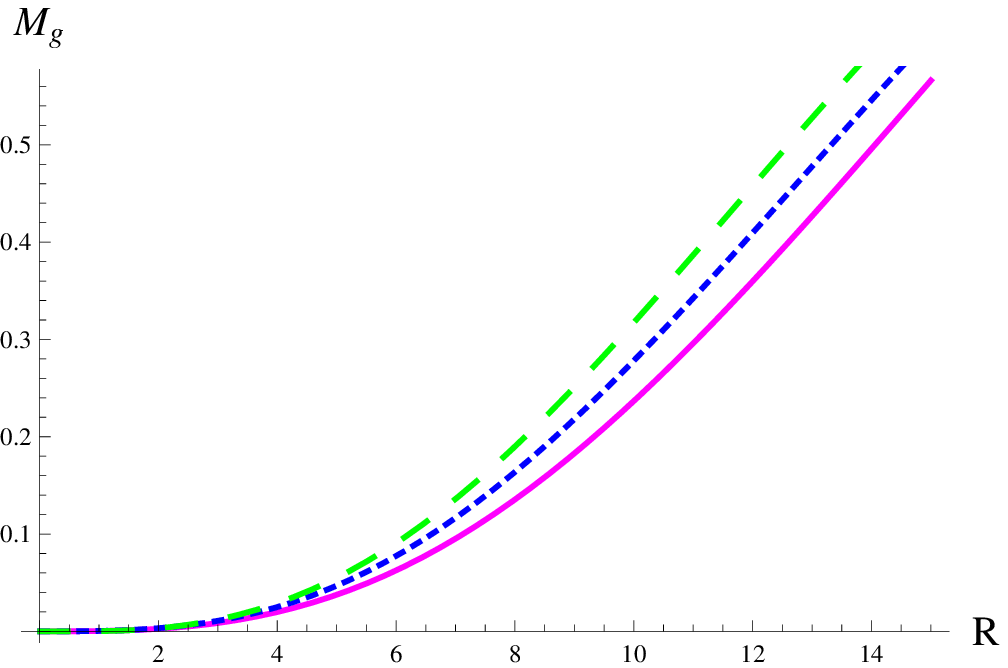,width=0.35\linewidth}\epsfig{file=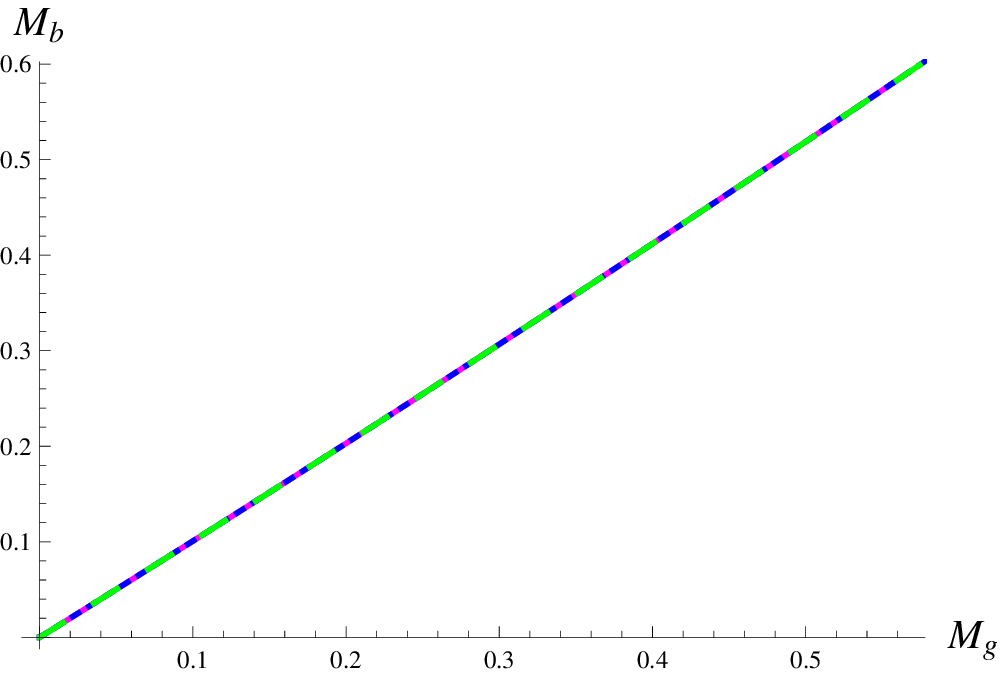,width=0.35\linewidth}
\caption{Plots of gravitational mass versus radius (left) and
baryonic mass versus gravitational mass (right) corresponding to
extended KB solution for case II.}
\end{figure}
\begin{figure}\center
\epsfig{file=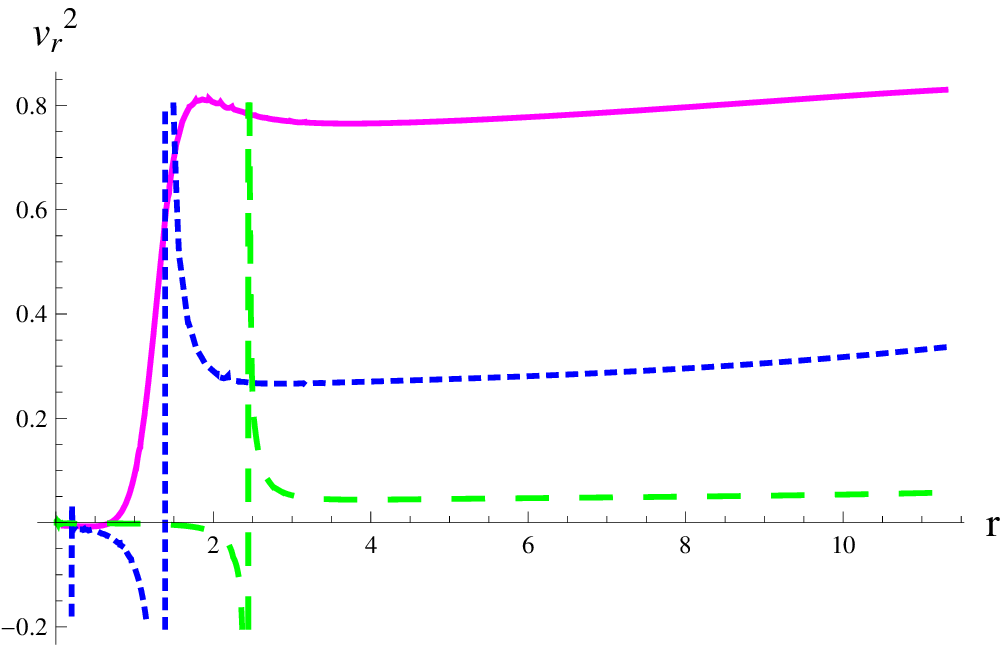,width=0.35\linewidth}\epsfig{file=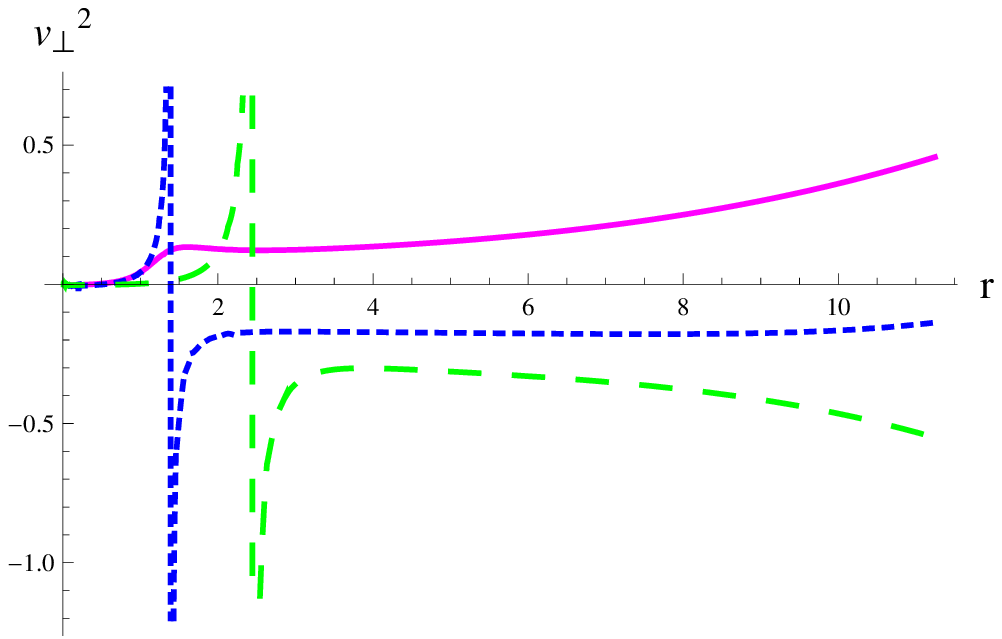,width=0.35\linewidth}\epsfig{file=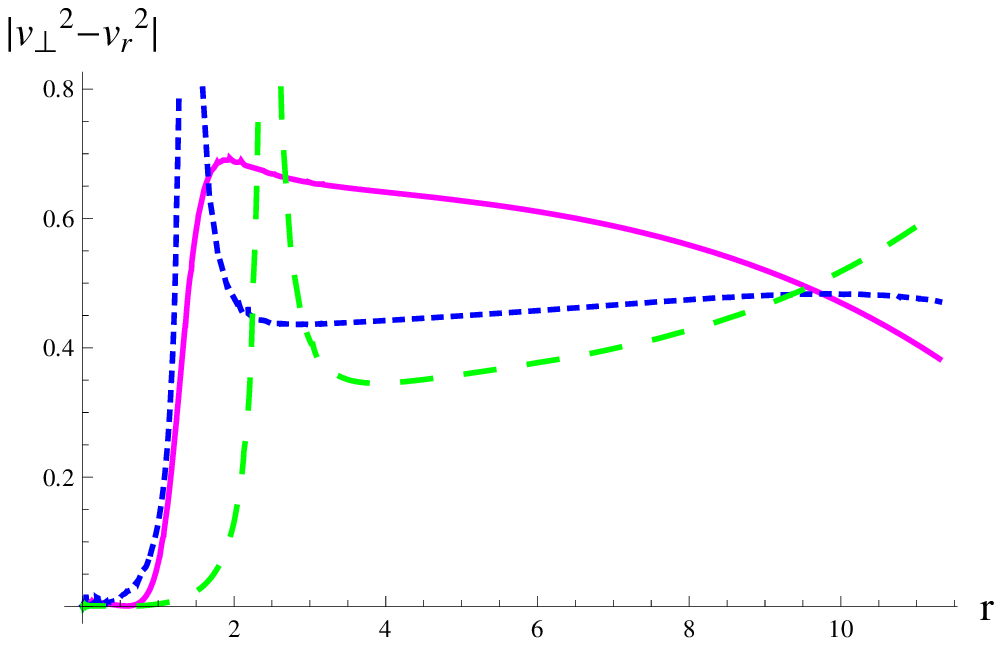,width=0.35\linewidth}
\caption{Plots of radial/tangential velocities and
$|v_\perp^2-v_r^2|$ corresponding to extended KB solution for case
II.}
\end{figure}

\section{Conclusions}

In this paper, we have formulated anisotropic solutions by
introducing a new source in the perfect fluid distribution in the
framework of SBD gravity. For this purpose, the field equations have
been decoupled into two sets via the EGD technique. To examine the
efficiency of this method, we have specified the first set by
considering metric coefficients of two isotropic solutions: Tolman
IV and KB. The scalar field has been obtained by solving the wave
equation numerically for $m_\Psi=0.01$. The number of unknown
variables in the anisotropic sector has been reduced through two
constraints on the additional source. Finally, we have inspected
physical properties of the constructed models through energy
conditions, compactness and redshift parameters for
$\alpha=0.2,~0.55,~0.9$. The obtained solutions have also been
checked for stability by employing two criteria: causality condition
and Herrera's cracking approach.

The first condition on $\Theta$-sector requires $\Theta_1^1$ to
mimic isotropic pressure. For the second constraint, we have
discussed two cases:
\begin{itemize}
\item A barotropic EoS relating $\Theta_0^0$ to $\Theta_1^1$;
\item A regularity condition on anisotropy on $\Theta^\gamma_\delta$ following Bowers-Lang constraint
\cite{35}.
\end{itemize}
In the first scenario, the deformation functions $f(r)$ and $g(r)$
have been calculated through EoS and mimic constraint, respectively.
The graphical analysis of state parameters of anisotropic Tolman IV
solution shows that energy density and pressure components follow
the accepted trend for $\alpha=0.2,~0.55$ whereas transverse
pressure monotonically increases for $\alpha=0.9$. However, the
tangential pressure corresponding to the anisotropic KB solution
decreases towards the boundary only for $\alpha=0.2$. The
anisotropic models represented by both solutions are viable as well
as obey Buchdahl limit for compactness and redshift. Moreover,
higher values of the decoupling parameter correspond to denser and
more compact stellar structures in both models. The extended Tolman
IV solution is stable for $\alpha=0.2,~0.55$ whereas anisotropic KB
solution is stable according to Herrera's cracking approach but
violates the causality condition $0\leq v_\perp^2\leq1$ for the
considered values of $\alpha$.

For case II, matter variables of the anisotropic version of Tolman
IV solution are positive and decrease monotonically for the chosen
values of $\alpha$. On the other hand, the energy density of
extended KB solution decreases monotonically only for $\alpha=0.2$
while for $\alpha=0.55,~0.9$, the density decreases after increasing
for some distance. For these values of the decoupling parameter,
tangential pressure increases towards the boundary. However, both
solutions satisfy energy conditions as well as the limits on
compactness and surface redshift. Moreover, increase in the
decoupling parameter leads to a decrease in the density and
compactness of both anisotropic models. Finally, the strength of the
repulsive force due to positive anisotropy increases with the
increase in $\alpha$ in all four solutions. The model corresponding
to extended Tolman IV solution is consistent with both stability
criteria whereas anisotropic KB solution is stable for $\alpha=0.2$
only. It is inferred that viability of the extended Tolman IV
solutions in GR \cite{43,37} is preserved in SBD gravity as well.
Moreover, the anisotropic analogues of KB solution obtained here are
viable for higher values of the decoupling parameter in contrast to
the extended KB solutions obtained through MGD technique in
\cite{27}. Thus, EGD method yields anisotropic solutions with
suitable physical properties. It is interesting to mention here that
all the results of GR can be retrieved for $\Psi=\text{constant}$
and $\omega_{BD}\rightarrow \infty$.

\vspace{0.25cm}

{\bf Acknowledgment}

\vspace{0.25cm}

This work has been supported by the \emph{Pakistan Academy of
Sciences Project}.


\vspace{0.25cm}

\begin{thebibliography}{40}

\bibitem{1} Schwarzschild, K.: Math. Phys. \textbf{189}(1916).

\bibitem{1'} Lemaitre, G.: Ann. Soc. Sci. Bruxells A
\textbf{53}(1993)51.

\bibitem{2} Ruderman, A.: Annu. Rev. Astron. Astrophys.
\textbf{10}(1972)427.

\bibitem{7} Herrera, L. and Santos, N.O.: Phys. Reports
\textbf{286}(1997)53.

\bibitem{8} Harko, T. and Mak, M.K.: Annalen Phys.
\textbf{11}(2002)3.

\bibitem{9'} Paul, B.C. and Deb, R.: Astrophys. Space Sci.
\textbf{354}(2014)421.

\bibitem{9} Murad, M.H.: Astrophys. Space Sci.
\textbf{20}(2016)361.

\bibitem{10} Ovalle, J.: Mod. Phys. Lett. A \textbf{23}(2008)3247.

\bibitem{11} Ovalle, J. and Linares, F.: Phys. Rev. D \textbf{88}(2013)104026.

\bibitem{13} Ovalle, J. et al.: Eur. Phys. J. C \textbf{78}(2018)122.

\bibitem{15} Gabbanelli, L., Rinc´on, A. and Rubio, C.: Eur. Phys. J. C ´
\textbf{78}(2018)370.

\bibitem{16} Estrada, M. and Tello-Ortiz, F.: Eur. Phys. J. C \textbf{133}(2018)453.

\bibitem{17} Sharif, M. and Sadiq, S.: Eur. Phys. J. C \textbf{78}(2018)410.

\bibitem{16a} Hensh, S. and Stuchlik, Z.: Eur. Phys. J. C \textbf{79}(2019)834.

\bibitem{17'} Sharif, M. and Ama-Tul-Mughani, Q.: Int. J. Geom. Methods Mod. Phys. \textbf{16}(2019)1950187;
Mod. Phys. Lett. A \textbf{35}(2020)2050091.

\bibitem{40} Casadio, R., Ovalle, J. and da Rocha, R.: Class.
Quantum Grav. \textbf{32}(2015)215020.

\bibitem{41} Ovalle, J.: Phys. Lett. B \textbf{788}(2019)213.

\bibitem{42} Contreras, E. and Bargue\~{n}o, P.: Class.
Quantum Grav. \textbf{36}(2019)215009.

\bibitem{43} Sharif, M. and Ama-Tul-Mughani, Q.: Ann. Phys.
\textbf{415}(2020)168122.

\bibitem{44} Sharif, M. and Ama-Tul-Mughani, Q.: Chinese J. Phys. \textbf{65}(2020)207.

\bibitem{28}  Sharif, M. and Saba, S.: Eur. Phys. J. C
\textbf{78}(2018)921; Sharif, M. and Waseem, A.: Ann. Phys.
\textbf{405}(2019)14; Chin. J. Phys. \textbf{60}(2019)426; Sharif,
M. and Saba, S., \emph{Extended Gravitational Decoupling Approach in
f($\mathcal{G}$) Gravity}, Int. J. Mod. Phys. D (to appear, 2020),
https://doi.org/10.1142/S0218271820500418.

\bibitem{18'} Dirac, P.A.M: Nature \textbf{139}(1937)323; Proc.
R. Soc. Lond. A \textbf{165}(1938)199.

\bibitem{18} Brans, C. and Dicke, R.H.: Phys. Rev. \textbf{124}(1961)3.

\bibitem{19} Weinberg, E.J.: Phys. Rev. D \textbf{40}(1989)3950.

\bibitem{20} Will, C.M.: Living Rev. Rel. \textbf{4}(2001)4.

\bibitem{21} Khoury, J. and Weltman, A.: Phys. Rev. D \textbf{69}(2004)044026.

\bibitem{23'} Perivolaropoulos, L.: Phys. Rev. D \textbf{81}(2010)047501.

\bibitem{22'} Buchdahl, H.A.: Int. J. Theor. Phys.
\textbf{6}(1972)407.

\bibitem{22} Sneddon, G.E. and McIntosh, C.B.G.: Aust. J. Phys.
\textbf{27}(1974)411.

\bibitem{23} Geroch, R.: J. Math. Phys. \textbf{12}(1971)918.

\bibitem{24} Bruckman, W.F. and Kazes, E.: Phys. Rev. D
\textbf{16}(1977)2.

\bibitem{25} Goswami, G.K.: J. Math. Phys. \textbf{19}(1978)442.

\bibitem{25'} Johri, V.B. and Goswami, G.K.: J. Math. Phys. \textbf{19}(1978)987;
ibid. \textbf{21}(1980)2269.

\bibitem{25a} Krori, K.D. and Bhattacharjee, D.R.: J. Math. Phys.
\textbf{23}(1982)1846.

\bibitem{26} Riazi, N. and Askari, H.R.: Mon. Not. R. Astron. Soc.
\textbf{261}(1993)229.

\bibitem{27} Sharif, M. and Majid, A.: Astrophys. Space Sci. \textbf{365}(2020)42.

\bibitem{55} Yazadjiev, S.S., Doneva, D.D. and Popchev, D.: Phys.
Rev. D \textbf{93}(2016)084038.

\bibitem{56} Ramazano\~{g}lu, F.M. and Pretorius, F.: Phys.
Rev. D \textbf{93}(2016)064005.

\bibitem{57} Doneva, D.D. and Yazadjiev, S.S.: J. Cosmol. Astropart.
Phys. \textbf{11}(2016)019.

\bibitem{58} Staykov, K.V. et al.: Eur. Phys. J. C
\textbf{78}(2018)586.

\bibitem{59} Popchev, D. et al.: Eur. Phys. J. C
\textbf{79}(2019)178.

\bibitem{30} Tolman, R.C.: Phys. Rev. \textbf{55}(1939)364.

\bibitem{29c} Delgaty, M.S.R. and Lake, K.: Comput. Phys. Commun. \textbf{115}(1998)395.

\bibitem{30a} Bhar, P., Singh, K.N. and Manna, T.: Astrophys. Space Sci.
\textbf{361}(2016)284; Banerjee, S.: Pramana-J. Phys.
\textbf{91}(2018)27.

\bibitem{30b} Krori, K.D. and Barua, J.: J. Phys. A: Math. Gen.
\textbf{8}(1975)4.

\bibitem{31'} Momeni, D. et al.: Int. J. Mod.
Phys. A \textbf{30}(2015)1550093; Zubair, M. and Abbas, G.:
Astrophys. Space Sci. \textbf{361}(2016)342.

\bibitem{34} Fujii, Y. and Maeda, K.: \emph{The Scalar-Tensor Theory of Gravitation}
(Cambridge University Press, 2003).

\bibitem{32} Buchdahl, H.A.: Phys. Rev. D \textbf{116}(1959)1027.

\bibitem{33} Ivanov, B.V.: Phys. Rev. D \textbf{65}(2002)104011.

\bibitem{51} Burrows, A. and Lattimer, J.M.: Astrophys. J. \textbf{307}(1986)178;
Lattimer, J.M. and Yahil, A.: Astrophys. J. \textbf{340}(1989)426.

\bibitem{52} Timmes, F.X., Woosley,, S.E. and Weaver, T.A.:
Astrophys. J. \textbf{457}(1996)834.

\bibitem{53} Abreu, H., Hernandez, H. and Nunez, L.A.: Class. Quantum Gravit. \textbf{24}(2007)4631.

\bibitem{54} Herrera, L.: Phys. Lett. A \textbf{165}(1992)206.

\bibitem{35} Bowers, R. and Liang, E.: Astrophys. J. \textbf{188}(1974)657.

\bibitem{36}  Cosenza, M. et al.: J. of Math. Phys.
\textbf{22}(1981)118.

\bibitem{37} Abell\'{a}n, G.: Eur. Phys. J. C \textbf{80}(2020)177

\bibitem{50} Contreras, E., Tello-Ortíz, F. and Maurya, S.K.:
arXiv:2002.12444.
\end{thebibliography}
\end{document}